\documentclass[pre,twocolumn,showpacs,preprintnumbers,10pt]{revtex4}

\usepackage{graphicx, epsfig, subfigure, color}
\usepackage{amsfonts,amssymb,amsmath,amsthm,latexsym,times,enumerate}
\usepackage[english]{babel}

\begin{document}

\title{Quantifying Causal Coupling Strength:\\ A Lag-specific Measure For Multivariate Time Series Related To Transfer Entropy}

\author{Jakob Runge$^{1,2)}$, Jobst Heitzig$^{1)}$, Norbert Marwan$^{1)}$, and J\"urgen Kurths$^{1,2,3)}$}
\affiliation{ $^{1)}$ Potsdam Institute for Climate Impact Research (PIK), 14473 Potsdam, Germany\\
$^{2)}$ Department of Physics, Humboldt University, 12489 Berlin, Germany\\
$^{3)}$ Institute for Complex Systems and Mathematical Biology, University of Aberdeen, Aberdeen AB24 3UE, United Kingdom
}

\date{\today}
\begin{abstract}
While it is an important problem to identify the existence of causal associations between two components of a multivariate time series, a topic addressed in [J. Runge, J. Heitzig, V. Petoukhov, and J. Kurths, \emph{Physical Review Letters} 108, 258701 (2012)], it is even more important to assess the strength of their association in a meaningful way. 
In the present article we focus on the problem of defining a meaningful coupling strength using information theoretic measures and demonstrate the short-comings of the well-known mutual information and transfer entropy.
Instead, we propose a certain time-delayed conditional mutual information, the \emph{momentary information transfer} (MIT), as a measure of association that is general, causal and lag-specific, reflects a well interpretable notion of coupling strength and is practically computable.
Rooted in information theory, MIT is general, in that it does not assume a certain model class underlying the process that generates the time series. 
As discussed in a previous paper [J. Runge, J. Heitzig, V. Petoukhov, and J. Kurths, \emph{Physical Review Letters}  108, 258701 (2012)], the general framework of graphical models makes MIT causal, in that it gives a non-zero value only to lagged components that are not independent conditional on the remaining process. Further, graphical models admit a low-dimensional formulation of conditions which is important for a reliable estimation of conditional mutual information and thus makes MIT practically computable. 
MIT is based on the fundamental concept of source entropy, which we utilize to yield a notion of coupling strength that is, compared to mutual information and transfer entropy, well interpretable, in that for many cases it solely depends on the interaction of the two components at a certain lag. In particular, MIT is thus in many cases able to exclude the misleading influence of autodependency within a process in an information-theoretic way. We formalize and prove this idea analytically and numerically for a general class of nonlinear stochastic processes and illustrate the potential of MIT on climatological data.

\end{abstract}


\maketitle

\section{Introduction}
Today's scientific world produces a vastly growing and technology-driven abundance of data across all research fields from observations of natural processes to economic data \cite{science2011}. To test or generate hypotheses on interdependencies between processes underlying the data, statistical measures of association are needed. 
Recently, Reshef \textit{et al.} \cite{Reshef2011} put forward two key demands such a measure should fulfill in the bivariate case: (1) \emph{generality}, i.e., the measure should not be restricted to certain types of associations like linear measures, and (2) \emph{equitability}, which means that the measure should reflect a certain heuristic notion of coupling strength, i.e., it should give similar scores to equally noisy dependencies. The latter is especially important for comparisons and ranking of the strength of dependencies.
In this article we generalize this idea to multivariate data as needed to reconstruct interaction networks in the fields of neuroscience, genetics, climate, ecology and many more. For the multivariate case we propose to add two more basic properties: (3) \emph{causality}, which means that the measure should give a non-zero value only to the dependency between lagged components of a multivariate process that are not independent conditional on the remaining process. (4) \emph{coupling strength autonomy}, implying that also for dependent components we seek for a causal notion of coupling strength that is well interpretable, in that it is uniquely determined by the interaction of the two components alone and in a way autonomous of their interaction with the remaining process. To understand this, consider a simple example: Suppose we have two interacting processes $X$ and $Y$ and a third process $Z$, that drives both of them. Then a bivariate measure of coupling strength between $X$ and $Y$ will be influenced by the common input of $Z$, while our demand is, that the measure should be autonomous of the interactions of $X$ and $Y$ with $Z$. In an experimental setting this corresponds to keeping $Z$ fixed and solely measuring the impact of a change in $X$ on $Y$ averaged over all realizations of $Z$. This property can be regarded as one ingredient of a multivariate extension of the equitability property.
Last, we also demand that the measure should be defined in a way that is \emph{practically computable}, in that the estimation does not, e.g., require somewhat arbitrary truncations like in the case of transfer entropy \cite{Schreiber2000}.
Due to these properties our approach can be used to reconstruct interaction networks where not only the links are causal, but are also meaningfully weighted and have the attribute of a coupling delay. This serves as an important feature in inferring physical mechanisms from interpreting interaction networks.
\begin{figure*}[t]
\includegraphics[width=2\columnwidth]{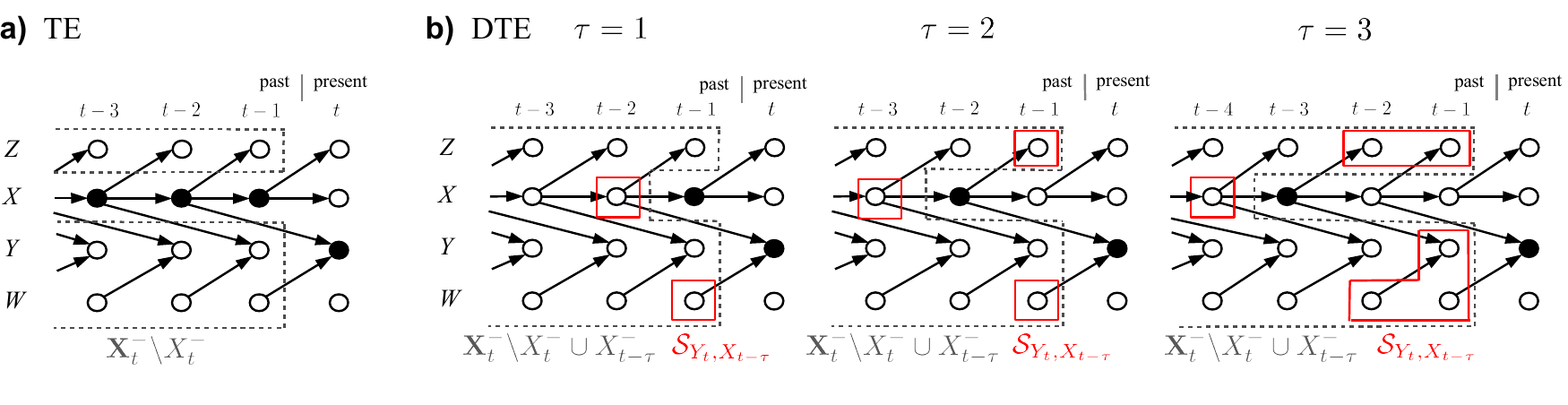}
\caption[]{(Color online) TE and DTE for a multivariate example process as given by Eq.~(\ref{eq:ana_model}) that will be analytically analyzed in Sect.~\ref{sect:analytics}. The time series graph is defined in Sect.~\ref{sect:graph}. (a) depicts the TE between the infinite past vector $X^-_t$ and $Y_t$ (black dots) conditioned on the remaining infinite past $\mathbf{X}_t^-{\setminus}{X_t^-}$ (gray dashed open box). (b) illustrates the first three summands of DTE given by Eq.~(\ref{eq:sum}). For the CMI between $X_{t-\tau}$ and $Y_t$ (black dots) only the finite set $\mathcal{S}_{Y_t,X_{t-\tau}}$ (red solid boxes) is needed to satisfy the Markov property (Eq.~(2) in \cite{Runge2012prl}). $\mathcal{S}_{Y_t,X_{t-\tau}} \subset \mathbf{X}_t^-{\setminus}{X_t^-}\cup X_{t-\tau}^-$ (gray dashed open box) must be chosen so that it \textit{separates} the remaining infinite conditions $(\mathbf{X}_t^-{\setminus}{X_t^-}\cup X_{t-\tau}^-){\setminus}\mathcal{S}_{Y_t,X_{t-\tau}}$ from $Y_t$ in the graph (for a formal definition of paths and separation see \cite{Eichler2011}). Since the separating sets depend on paths between $\mathbf{X}_t^-{\setminus}{X_t^-}\cup X_{t-\tau}^-$ and $Y_t$, they can only be determined after the time series graph has been estimated.
}
\label{fig:graph_te_dte}
\end{figure*}

The first requirement, generality, is fulfilled by any information theoretic measure like mutual information (MI) and conditional mutual information (CMI) \cite{Cover2006}. These measures also fulfill the axioms for dependency measures proposed in \cite{Schweizer1981}. Additionally to generality, the authors in \cite{Reshef2011} demonstrate that their algorithmically motivated maximal information coefficient fulfills the property of equitability. However, apart from issues with statistical power \cite{Gorfine2012}, a crucial drawback of their measure is, that it is not clear how to extend it to the multivariate case. There are few works considering a concept of coupling strength in the multivariate context of causality. In \cite{Jachan2009a,Schelter2009} this problem is approached in the linear framework of partial directed coherence and in \cite{Chen2004,Marinazzo2008} using the less restricted, yet still model-based, concept of Granger causality, all sharing the problem that the model might be misspecified. Transfer entropy (TE) \cite{Schreiber2000} is the information-theoretic analogue of Granger causality \cite{Barnett2009} and the issue of arbitrary truncations has been addressed in \cite{Faes2011} and in our previous article \cite{Runge2012prl}. Still the problem with TE is that it is not lag-specific which can lead to false interpretations like in the case of feedbacks \cite{ay2008information} and, as we will demonstrate analytically and numerically in this article, it is \emph{not} uniquely determined by the interaction of the two components alone and depends on misleading effects of, e.g., autodependency and the interaction with other processes. In essence, it does not fulfill the proposed property of coupling strength autonomy. In \cite{Janzing2012} another information-theoretic approach, based on a different set of postulates, is discussed.

Our approach to a measure of a causal coupling strength is based on the fundamental concept of \emph{source entropy} \cite{Shannon1963} and for the special case of bivariate ordinal pattern time series the \textit{momentary information transfer} (MIT) has been introduced recently in \cite{Pompe2011}. 
In this article we utilize the concept of graphical models to mathematically formalize and generalize MIT to the multivariate case. We demonstrate that MIT is practically computable and fulfills the properties of generality, causality and coupling strength autonomy, while the more complex property of equitability will only partially be addressed here. 

The determination of a causal coupling strength in our approach is a two-step process. In the first step the graphical model is estimated as detailed in \cite{Runge2012prl} which determines the existence or absence of a link and thus of a causality between lagged components of the multivariate process. The second step -- the main topic of the present paper -- is the estimation of MIT as a meaningful weight for every existing link in the graph.

The article is organized as follows. In Sect.~\ref{sect:te} we define and review TE and the \emph{decomposed transfer entropy} introduced in \cite{Runge2012prl}. In Sect.~\ref{sect:graph} we introduce the important concept of graphical models and in Sect.~\ref{sect:mit} we define MIT and related measures. All of these measures are compared analytically (Sect.~\ref{sect:analytics}), leading to the coupling strength autonomy theorem (Sect.~\ref{sect:coupling_independence}), and numerically (Sect.~\ref{sect:numerics}). Finally, we discuss limitations (Sect.~\ref{sect:limitations}) and provide an application to climatological data that shows the potential of our approach (Sect.~\ref{sect:application}). The appendices provide proofs and further discussions. 


\section{Transfer Entropy and the curse of dimensionality} \label{sect:te}
Before introducing MIT, we will discuss the well-known TE and its short-comings. We will focus on multivariate time series generated by discrete-time stochastic processes and use the following notation: Given a stationary multivariate discrete-time stochastic process $\mathbf{X}$, we denote its uni- or multivariate subprocesses $X,Y,Z,W,\ldots$ and the random variables at time $t$ as $\mathbf{X}_t,X_t,\ldots$. Their \emph{pasts} are defined as $\mathbf{X}_t^-=(\mathbf{X}_{t-1},\mathbf{X}_{t-2},\ldots)$ and $X_t^-=(X_{t-1},X_{t-2},\ldots)$. For convenience, we will often treat $\mathbf{X}$, $\mathbf{X}_t$, $\mathbf{X}_t^-$, and $X^-_t$ as sets of random variables, so that, e.g., $X_t^-$ can be considered a subset of $\mathbf{X}_t^-$. 
Now the TE [see Fig.~\ref{fig:graph_te_dte}(a)]
\begin{align} \label{eq:def_te}
I^{\rm TE}_{X {\to} Y} &\equiv I(X_t^-;Y_t\,|\,\mathbf{X}_t^- {\setminus} X_t^-)
\end{align}
is the reduction in uncertainty about $Y_t$ when learning the past of $X_t$, if the rest of the past of $\mathbf{X}_t$, given by $\mathbf{X}_t^- {\setminus} X_t^-$, is already known (where ``$\setminus$'' denotes the subtraction of a set). Note that, because of the assumed stationarity, $I^{\rm TE}_{X\to Y}$ is independent of $t$. TE measures the aggregated influence of $X$ at all past lags and is not lag-specific. The definition of TE leads to the problem that infinite-dimensional densities have to be estimated, which is commonly called the ``curse of dimensionality''. In the usual naive estimation of TE the infinite vectors are simply truncated at some $\tau_{\max}$ leading to
\begin{align} \label{eq:te_trunc}
I^{{\rm TE}, \tau_{\max}}_{X {\to} Y} &\equiv I(X_t^{(t-1,\ldots,t-\tau_{\max})};Y_t\,|\,\mathbf{X}_t^{(t-1,\ldots,t-\tau_{\max})} {\setminus} X_t^-).
\end{align}
where $X_t^{(t-1,\ldots,t-\tau_{\max})}=(X_{t-1},\,\ldots,X_{t-\tau_{\max}})$ (correspondingly for $\mathbf{X}$) and $\tau_{\max}$ has to be chosen at least as large as the maximal coupling delay between $X$ and $Y$, which can lead to very large dimensions. In our numerical experiments we will demonstrate that the choice of a truncation lag $\tau_{\max}$, which affects the estimation dimension via $D=N\cdot\tau_{\max}+1$ (where $N$ is the number of processes), has a strong influence on the value of TE and affects the reliability of causal inference. This is a huge disadvantage because the coupling delay should not have an influence on the measured coupling strength.

In \cite{Runge2012prl} the problem of high dimensionality is overcome by utilizing the concept of graphical models that will be introduced in the next section. In this framework a decomposed transfer entropy (DTE) is derived that enables an estimation using finite vectors
\begin{align} \label{eq:sum}
 I^{\rm TE}_{X{\to} Y} \approx I^{\rm DTE}_{X{\to} Y} \equiv \sum_{\tau=1}^{\tau^\star} I(X_{t-\tau};Y_t\,|\, \mathcal{S}_{Y_t,X_{t-\tau}})
\end{align}
for a certain \textit{finite} set $\mathcal{S}_{Y_t,X_{t-\tau}} \subset \mathbf{X}_t^-{\setminus}{X_t^-}\cup X_{t-\tau}^-$ [see Fig.~\ref{fig:graph_te_dte}(b)] and with $\tau^\star$ chosen as the smallest $\tau$ for which the estimated remainder is smaller than some given threshold. Another approach to find a truncation is described in \cite{Faes2011}. While thereby the somewhat arbitrary truncation lag $\tau_{\max}$ is avoided and the estimation dimension is drastically reduced, it can still be quite high (in the still rather simple model example of \cite{Runge2012prl} the maximum dimension was 24). 

The summands in Eq.~(\ref{eq:sum}) can be seen as the contributions of different lags to TE, but should not be interpreted as lag-specific causal contributions because they can be non-zero also for lags $\tau$ for which there is no link in the graph. Finally, apart from the issue of high dimensionality and lag-specific causality, we will demonstrate in Sect.~\ref{sect:analytics} that TE or DTE also do not fulfill the proposed coupling strength autonomy property. In the next section we introduce the important concept of graphical models from which we derive MIT and related measures.

\begin{figure}[t]
\includegraphics[width=\columnwidth]{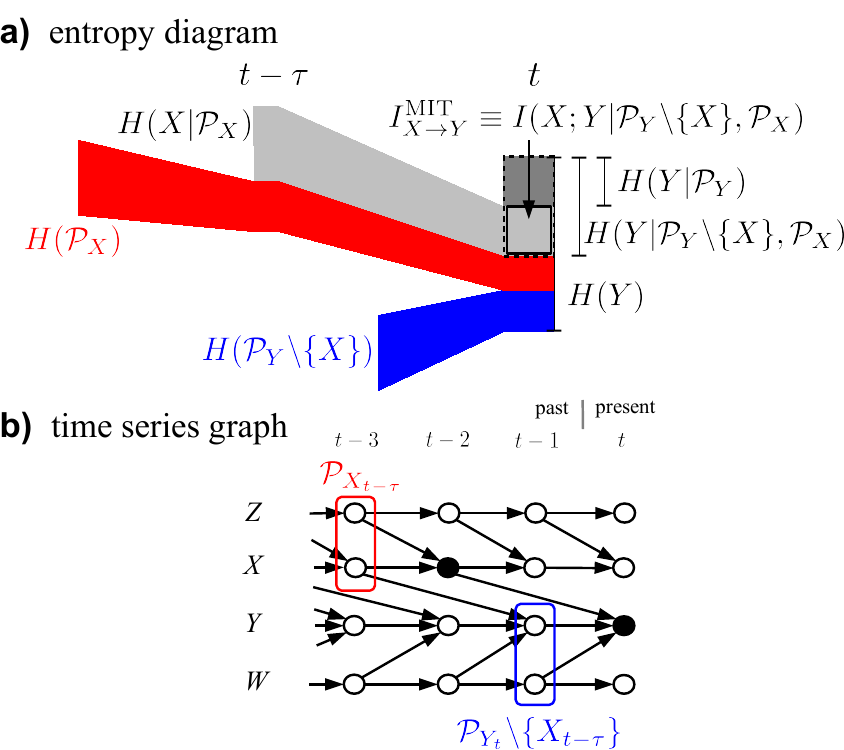}
\caption[]{(Color online) (a) Venn diagram that depicts the entropy $H(Y)$ at time $t$ (omitting $t$ and $\tau$ in the labels) as a segmented column bar. It is composed of the source entropy $H(Y|\mathcal{P}_Y)$ (dark gray shaded) and parts of the source entropy $H(X|\mathcal{P}_X)$ (light gray shaded), the entropy $H(\mathcal{P}_X)$ of the parents of $X$ (red), and the entropy $H(\mathcal{P}_Y{\setminus}\{X_{t-\tau}\})$ of the remaining parents of $Y$ (blue).
Our CMI $I^{\rm MIT}_{X{\to}Y}$ (solid framed segment) is the difference between the entropy $H(Y| \mathcal{P}_{Y} {\setminus}\{X\},\mathcal{P}_{X} )$ (dashed segment) that includes transfer from $X$ and the source entropy of $Y$ that excludes it.
(b) shows an example of a time series graph (see definition in text) corresponding to Eq.~(\ref{eq:model}) that makes the intuitive entropy picture operational. In this graph MIT is the CMI between $X_{t-\tau}$ at $\tau=2$ and $Y_t$ (marked by the black dots) conditioned on the parents $\mathcal{P}_{X_{t-\tau}}$ (red) and $\mathcal{P}_{Y_t} {\setminus} \{X_{t-\tau}\}$ (blue).
}
\label{fig:entropy_flow}
\end{figure}

\section{Graphical Models and Causality} \label{sect:graph}
In the graphical model approach \cite{lauritzen1996graphical,Dahlhaus2000,Eichler2011} the conditional independence properties of a multivariate process are visualized in a graph, in our case a time series graph. This graph thus encodes the lag-specific causality with respect to the observed process.
As depicted in Figs.~\ref{fig:graph_te_dte} and \ref{fig:entropy_flow}(b), each node in that graph represents a single random variable, i.e., a subprocess, at a certain time $t$. Nodes $X_{t-\tau}$ and $Y_t$ are connected by a directed link ``$X_{t-\tau}~\to~Y_t$'' pointing forward in time if and only if $\tau>0$ and 
\begin{align}  \label{eq:def_graph}
   I^{\rm LINK}_{X\to Y}(\tau) &\equiv I(X_{t-\tau}; Y_t| \mathbf{X}_t^-\setminus \{X_{t-\tau}\}) > 0,
\end{align}
i.e., if they are not independent conditionally on the past of the whole process, which implies a lag-specific causality with respect to $\mathbf{X}$.
If $Y\neq X$ we say that the link ``$X_{t-\tau} \to Y_t$'' represents a \textit{coupling at lag} $\tau$, while for $Y=X$ it represents an \textit{autodependency at lag} $\tau$. Nodes $X_t$ and $Y_t$ are connected by an undirected contemporaneous link (visualized by a line) \cite{Eichler2011} if and only if
\begin{align} \label{eq:def_graph_contemp}
I^{\rm LINK}_{X\-- Y} &\equiv I(X_t; Y_t\,|\, \mathbf{X}_{t+1}^- {\setminus} \{X_t,Y_t\}) > 0
\end{align}
where also the contemporaneous present $\mathbf{X}_t{\setminus}\{X_t,Y_t\}$ is included in the condition. In the case of a multivariate autoregressive process as defined later in Eq.~(\ref{eq:var}), this definition corresponds to non-zero entries in the \emph{inverse} covariance matrix of the innovations $\mathbf{\varepsilon}$.
Note that stationarity implies that ``$X_{t-\tau} \to Y_t$'' whenever ``$X_{t'-\tau}\to Y_{t'}$'' for any $t'$.

Like TE, the CMIs given by Eq.~(\ref{eq:def_graph}) and (\ref{eq:def_graph_contemp}) involve infinite-dimensional vectors and can thus not be directly computed, but only involving truncations. As shown in Sect.~\ref{sect:numerics}, this measure therefore suffers from the problem of high dimensionality and also theoretically does not fulfill the coupling strength autonomy property as analyzed in Sect.~\ref{sect:analytics}. 

On the other hand, one can exploit the Markov property and use the finite set of \textit{parents} defined as
\begin{align}
\mathcal{P}_{Y_t} \equiv \{Z_{t-\tau}:~ Z\in \mathbf{X},~\tau>0,~Z_{t-\tau}\to Y_t\}
\end{align}
of $Y_t$ [blue box in Fig.~\ref{fig:entropy_flow}(b)] which separate $Y_t$ from the past of the whole process $\mathbf{X}_t^- {\setminus} \mathcal{P}_{Y_t}$. The parents of all subprocesses in $\mathbf{X}$ together with the contemporaneous links comprise the time series graph. In \cite{Runge2012prl} an algorithm for the estimation of these time series graphs by iteratively inferring the parents is introduced. In the Supplementary Material of \cite{Runge2012prl} we also describe a suitable shuffle test and a detailed numerical study on the detection and false positive rates of the algorithm. The Markov properties hold for models sufficing the very general condition\,(S) in \cite{Eichler2011}.

The determination of a causal coupling strength now is a two-step procedure. In the first step the time series graph is estimated as detailed in \cite{Runge2012prl} which determines the existence or absence of a link and thus of a causality between lagged components of $\mathbf{X}$. The second step is the determination of a meaningful weight for every existing link in the graph.
The MIT introduced in the next section is intended to serve this aim by attributing a well interpretable coupling strength solely to the inferred links of the time series graph.

\section{Momentary information transfer and source entropy} \label{sect:mit}
The parents of a subprocess $Y$ at a certain time $t$ are key to understand the underlying concept of source entropy.
Each univariate subprocess $X$ of a stationary multivariate discrete-time stochastic process $\mathbf{X}$ will at each time $t$ yield a realization $x_t$. 
The entropy of $X_t$ measures the uncertainty about $x_t$ before its observation, and it will in general be reduced if a realization of the parents $\mathcal{P}_{X_t} \subset \mathbf{X}_t^-$ is known. But for a non-deterministic process, and most real data will at least contain some random noise, there will always be some ``surprise'' left when observing $x_t$. This surprise gives us information and the expected information is called the source entropy $H(X_t|\mathcal{P}_{X_t})$ of $X$. 
Now the MIT between $X$ at some lagged time $t-\tau$ in the past and $Y$ at time $t$ is the CMI that measures the part of source entropy of $Y$ that is shared with the source entropy of $X$:
\begin{align} \label{eq:def_mit}
 I^{\rm MIT}_{X{\to}Y}(\tau) &\equiv I(X_{t-\tau};Y_t| \mathcal{P}_{Y_t} {\setminus}\{X_{t-\tau}\},\mathcal{P}_{X_{t-\tau}} )  \nonumber \\
 &= H(Y_t| \mathcal{P}_{Y_t} {\setminus}\{X_{t-\tau}\},\mathcal{P}_{X_{t-\tau}} ) - H(Y_t|\mathcal{P}_{Y_t}).
\end{align}
This approach of ``isolating source entropies'' is sketched in a Venn diagram in Fig.~\ref{fig:entropy_flow}(a). 
The attribute \textit{momentary} \cite{Pompe2011} is used because MIT measures the information of the ``moment'' $t-\tau$ in $X$ that is transferred to $Y_t$. This ``momentariness'' is closely related to the property of coupling strength autonomy as we will show in the next sections. Similarly to the definition of contemporaneous links in Eq.~(\ref{eq:def_graph_contemp}), we can also define a contemporaneous MIT
\begin{align} \label{eq:def_mit_contemp}
 I^{\rm MIT}_{X{\--}Y} &\equiv I(X_{t};Y_t| \mathcal{P}_{Y_t},\mathcal{P}_{X_{t}},\mathcal{N}_{X_t}{\setminus}\{Y_t\},\mathcal{N}_{Y_t}{\setminus}\{X_t\},\nonumber\\
&~~~~~~~~~~~~~~~~~~~~~\mathcal{P}(\mathcal{N}_{X_t}{\setminus}\{Y_t\}),\mathcal{P}(\mathcal{N}_{Y_t}{\setminus}\{X_t\}) )  
\end{align}
where $\mathcal{N}$ denotes the contemporaneous neighbors given by 
\begin{align}
\mathcal{N}_{Y_t} &\equiv \{X_t:X\in\mathbf{X}, X_t {\--} Y_t\}
\end{align}
and correspondingly for $X$ and their parents. Due to Markov properties the contemporaneous MIT is equivalent to the formula defining contemporaneous links Eq.~(\ref{eq:def_graph_contemp}). This is, however, not the case for the lagged MIT.
Like any (C)MI, MIT is sensitive to any kind of statistical association and therefore guarantees the property of generality. Because MIT uses the parents $\mathcal{P}_{Y_t}$ as conditions, it also fulfills the property of lag-specific causality in that it is non-zero only for lagged processes that are not independent conditional on $\mathbf{X}^-_t$.

As related measures, we can also choose either one of the parents as a condition, which -- dropping the attribute ``momentary'' -- leads to the \emph{information transfers} ITY and ITX
\begin{align} 
I_{X \to Y}^{\rm ITY}(\tau) &\equiv I(X_{t-\tau};Y_t|\mathcal{P}_{Y_t}{\setminus}\{X_{t-\tau}\}) \label{eq:def_py}, \\
I_{X \to Y}^{\rm ITX}(\tau) &\equiv I(X_{t-\tau};Y_t|\mathcal{P}_{X_{t-\tau}})  \label{eq:def_px}.
\end{align}
ITY isolates only the source entropy of $Y$. Like MIT it is non-zero only for dependent nodes (and therefore fulfills the properties of generality and causality) and used in the algorithm to estimate the time series graph \cite{Runge2012prl}. ITX measures the part of source entropy in $X_{t{-}\tau}$ that reaches $Y_t$ on any path and is, thus, not a causal measure, yet in many situations we might only be interested in the effect of $X$ on $Y$, no matter how this influence is mediated. 
For $\tau>0$ these three CMIs are related by the inequality
\begin{align} \label{eq:inequality}
 I_{X \to Y}^{\rm ITX}(\tau) \leq I^{\rm MIT}_{X{\to}Y}(\tau) \leq  I_{X \to Y}^{\rm ITY}(\tau),
\end{align}
which holds under the ``no sidepath''-constraint as specified in Sect.~\ref{sect:coupling_independence}. The proof is given in the appendix.
The very definition of MIT, ITY and ITX already leads to a low-dimensional estimation problem without arbitrary truncation parameters. Further, the underlying theory of time series graphs allows for an analytical evaluation of the properties of these measures as we will demonstrate in the following section. See \footnote{A \emph{Python}-script to estimate the time series graph, MIT and related measures can be obtained from the website {\tt http://tocsy.pik-potsdam.de/tigramite.php}.} for software to compute the time series graph, MIT and related measures.

To clarify, each of the CMIs introduced in the preceding sections are intended to measure a different aspect of the coupling between $X$ and $Y$. In the following analytical analysis of simple models we will discuss the interpretability of the different measures.

\section{Analytical comparison} \label{sect:analytics}   
To motivate our choice of a measure of coupling strength and to clarify the important coupling strength autonomy property, we discuss an analytically tractable model of a multivariate Gaussian process:
\begin{align} \label{eq:ana_model}
Z_t &= c_{XZ} X_{t-1} + \eta^Z_t \nonumber \\
X_t &= a_{X} X_{t-1} + \eta^X_t \nonumber \\
Y_t &= c_{XY} X_{t-2} + c_{WY} W_{t-1} + \eta^Y_t \nonumber\\
W_t &= \eta^W_t
\end{align}
with independent Gaussian white noise processes $\eta^{\cdot}_t$ with variances $\sigma^2_{\cdot}$. The corresponding time series graph is depicted in Fig.~\ref{fig:graph_te_dte} and the parents are $\mathcal{P}_{Y_t}=\{X_{t-2},W_{t-1}\}$ and $\mathcal{P}_{X_{t-2}}=\{X_{t-3}\}$. 
Generally, the conditional entropy $H(Y|\mathbf{Z})$ of a $D_Y$-dimensional Gaussian process $Y$ conditional on a (possibly multivariate) process $\mathbf{Z}$ is given by
\begin{align} \label{eq:cond_ent}
H(Y|\mathbf{Z}) &= \frac{1}{2} \ln \left( (2 \pi e)^{D_Y} \frac{|\Gamma_{Y \mathbf{Z}}|}{|\Gamma_{\mathbf{Z}}|} \right) 
\end{align}
where $|\Gamma_{Y \mathbf{Z}}|$ is the determinant of the covariance matrix of $(Y, \mathbf{Z})$. In our case $Y$ is univariate and thus $D_Y=1$.
The variances and covariances needed to evaluate the determinants and detailed derivations for the following formulas are given in the appendix.

First, we analyze TE given by Eq.~(\ref{eq:def_te}). TE can be written as the difference of conditional entropies
\begin{align}
I^{\rm TE}_{X {\to} Y} &= H(Y_t\,|\,\mathbf{X}_t^- {\setminus} X_t^-) - H(Y_t\,|\,\mathbf{X}_t^-),
\end{align}
where the latter entropy, conditioned on the whole infinite past, is actually the source entropy of $Y$ and can be much easier computed by exploiting the Markov property
\begin{align}
H(Y_t\,|\,\mathbf{X}_t^-) &= H(Y_t|\mathcal{P}_{Y_t}),
\end{align}
which yields, using Eq.~(\ref{eq:cond_ent}),
\begin{align}
H(Y_t|\mathcal{P}_{Y_t})&= \frac{1}{2} \ln \left( 2 \pi e \frac{|\Gamma_{Y_t X_{t-2}W_{t-1}}|}{|\Gamma_{X_{t-2},W_{t-1}}|} \right) \nonumber \\
                        &= \frac{1}{2} \ln \left( 2 \pi e \sigma_Y^2 \right).  \label{eq:ana_model_source}
\end{align}
The source entropy of $Y$ is therefore given by the entropy of the innovation term $\eta^Y$.
In the first entropy term, on the other hand, the infinite vector cannot be treated that easily and we have to evaluate the determinants of infinite dimensional matrices in
\begin{align}
H(Y_t|Y^-_t,W^-_t,Z^-_t) &= \frac{1}{2} \ln \left( 2 \pi e \frac{|\Gamma_{Y_t Y^-_t W^-_t Z^-_t}|}{|\Gamma_{Y^-_t W^-_t Z^-_t}|} \right).
\end{align}
However, for the special case of $c_{XZ}=c_{WY}=0$, i.e., no input processes apart from the autodependency in $X$, the quotient of these matrices can be simplified to the quotient of infinite Toeplitz matrices. As shown in the appendix, we can then apply Szeg\"o's theorem \cite{szegoe,boettcher2006} and get
\begin{align}
I^{\rm TE}_{X\to Y} &\stackrel{c_{XZ}=c_{WY}=0}{=}\frac{1}{2}\ln \left( 1 + \frac{(c_{XY}^2 \sigma_X^2)/(1{-}a_X^2)}{\sigma_Y^2} \right).
\end{align}
Another tractable case is $a_X=0$ for which the blocks of the covariance matrix $\Gamma_{Y_t Y^-_t W^-_t Z^-_t}$ become diagonal and 
\begin{align}
I^{\rm TE}_{X\to Y} &\stackrel{a_X=0}{=} \frac{1}{2}\ln \left( 1 + \frac{c_{XY}^2 \sigma_X^2 \sigma_Z^2}{\sigma_Y^2(c_{XZ}^2\sigma_X^2+\sigma_Z^2)} \right).
\end{align}
Thus, in the first case the value of TE for our model depends on the autodependency coefficient and in the second case on the coupling coefficient and variance of $Z$. But why should a measure of coupling strength between $X$ and $Y$ depend on internal dynamics of $X$ and, even more so, on the interaction of $X$ with another process $Z$? While it can be information-theoretically explained, it seems rather unintuitive for a measure of coupling strength between $X$ and $Y$.

Next, we compute the CMI $I^{\rm LINK}_{X\to Y}$ that defines links in a time series graph. Writing Eq.~(\ref{eq:def_graph}) for $\tau=2$ as a difference of conditional entropies, the second term is again the source entropy as given by Eq.~(\ref{eq:ana_model_source}) and in this case also the first entropy can be simplified using the Markov property
\begin{align}
H(Y_t\,|\,\mathbf{X}_t^-{\setminus}X_{t-2}) &= H(Y_t\,|\,\mathbf{X}_t^{(\mathbf{X}_{t-1},\ldots,\mathbf{X}_{t-3})}{\setminus}\{X_{t-2}\})
\end{align}
to arrive at a finite covariance matrix from which a lengthy computation yields
\begin{align} \label{eq:link_anamodel}
I^{\rm LINK}_{X\to Y} &= \frac{1}{2}\ln \left( 1 + \frac{c_{XY}^2 \sigma_X^2 \sigma_Z^2}{\sigma_Y^2(c_{XZ}^2\sigma_X^2+(1{+}a_X^2)\sigma_Z^2)} \right).
\end{align}
Again, also this measure of coupling strength depends on the coefficients belonging to other coupling and autodependency links. 

We now turn to the measures that solely use the parents as conditions which has the analytical and numerical advantage of low dimensional computations. The resulting expressions for the CMI with no conditions, i.e., the mutual information (MI), and for either one of the parents as a condition for $\tau=2$ are
\begin{align} 
I^{\rm MI}_{X\to Y} &= \frac{1}{2}\ln \left( 1 + \frac{(c_{XY}^2 \sigma_X^2)/(1{-}a_X^2)}{c_{WY}^2 \sigma_W^2+\sigma_Y^2} \right), \label{eq:mi}\\
I^{\rm ITY}_{X\to Y} &= \frac{1}{2}\ln \left( 1 + \frac{(c_{XY}^2 \sigma_X^2)/(1{-}a_X^2)}{\sigma_Y^2} \right), \label{eq:py} \\
I^{\rm ITX}_{X\to Y} &= \frac{1}{2}\ln \left( 1 + \frac{c_{XY}^2\sigma_X^2}{c_{WY}^2 \sigma_W^2+\sigma_Y^2} \right). \label{eq:px}
\end{align}
Thus MI depends on the coefficients and variances of the input processes, while ITX and ITY still depend at least on the coefficient and variance of the process that is not conditioned on. Contrary to TE and LINK though, neither of the three measures depends on the interaction with $Z$. 
In our model the inputs to $X$ and $Y$, i.e., the autodependency with $X_{t-3}$ and the external input from $W_{t-1}$, are independent which makes the formulas much simpler. 

Finally, the MIT for $\tau=2$ is
\begin{align} \label{eq:mit}
I^{\rm MIT}_{X\to Y} &= \frac{1}{2}\ln \left( 1 + \frac{c_{XY}^2\sigma_X^2}{\sigma_Y^2} \right),
\end{align}
which solely depends on the model coefficients that govern the source entropies, i.e., the variances $\sigma^2_X,\,\sigma^2_Y$, and the coupling coefficient $c_{XY}$.

This equation can be proven to hold for arbitrary multivariate linear autoregressive processes under the ``no sidepath''-constraint specified in the next section. More generally, for a class of additive models MIT depends only on the coupling coefficient $c_{XY}$ and the source variances of $\eta^X$ and $\eta^Y$ as will be proven in the coupling strength autonomy theorem in the next section. 

\begin{figure}[t]
\includegraphics[width=\columnwidth]{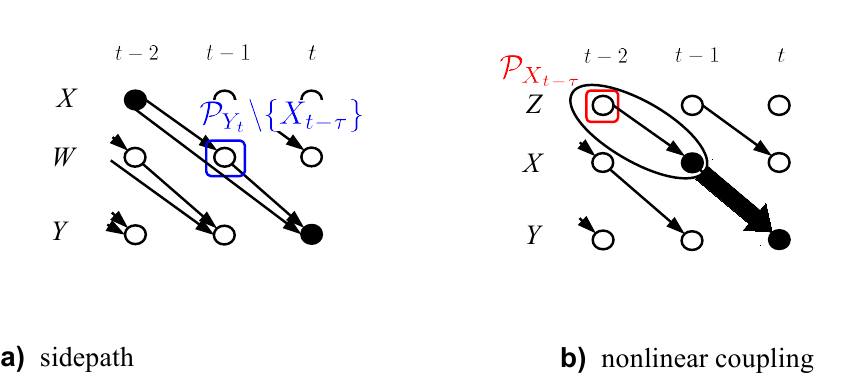}
\caption[]{(Color online) Two examples of couplings that cannot be related to one single coefficient $c_{XY}$. Black dots mark $X_{t-\tau}$ and $Y_t$, the red and blue boxes their parents.  (a) A sidepath, i.e., if there exists a path from $X_{t-2}$ to some parent of $Y_t$. Then the coupling cannot be related to one single link, but additionally to the path via $W_{t-1}$. (b) Visualization of a nonlinear coupling between $X_{t-1}$ and $Y_t$. In this case the entropies of $X_{t-1}$ and its parents ``mix'' and the coupling should be considered as emanating from $(X_{t-1},\mathcal{P}_{X_{t-1}})$ rather than $X_{t-1}$ alone.}
\label{fig:sidepath_nonlinear}
\end{figure}

But can a coupling strength always be associated with only one coupling coefficient $c_{XY}$? In the following -- still linear -- example model visualized in Fig.~\ref{fig:sidepath_nonlinear}(a)  this is not the case: 
\begin{align} \label{eq:ana_model_sidepath}
X_t &= \eta^X_{t} \nonumber  \\
W_t &= c_{XW} X_{t-1} + \eta^W_t  \nonumber  \\
Y_t &= c_{XY} X_{t-2} + c_{WY} W_{t-1} + \eta^Y_t 
\end{align}
where the influence of $X_{t-2}$ on $Y_t$ has two paths: One via the direct coupling link ``$X_{t-2}\to~Y_t$'' and one via the path ``$X_{t-2}\to W_{t-1}\to Y_t$'' such that we can rewrite 
\begin{align}
Y_t &= c_{XY} X_{t-2} + c_{WY} (c_{XW} X_{t-2} + \eta^W_{t-1}) + \eta^Y_t,
\end{align}
from which we see, that the coupling cannot be unambiguously related to one coefficient. Here, MIT at $\tau=2$ is
\begin{align} \label{eq:ana_mit_sidepath}
I^{\rm MIT}_{X\to Y}&=\frac{1}{2} \ln \left( 1 + \frac{c_{XY}^2 \sigma_X^2 \sigma_{W}^2}{\sigma_Y^2(c_{XW}^2 \sigma_X^2 +\sigma^2_{W})}  \right),
\end{align}
and depends not only on $c_{XY}$, but also on the coefficient $c_{XW}$ of the link ``$X_{t-2}\to~W_{t-1}$'', and on the variance of $W$.
In this case it might be more appropriate to ``leave open'' both paths and exclude $W_{t-1}$ from the conditions which -- only in this case -- reduces the modified MIT to the MI
\begin{align}
I(X_{t-2}; Y_t) &=\frac{1}{2} \ln \left(1+\frac{(c_{XY}+c_{XW} c_{WY})^2 \sigma_X^2}{c_{WY}^2 \sigma_W^2+\sigma_Y^2}\right).
\end{align}
Here the sum $c_{XY}+c_{XW} c_{WY}$ is the covariance along both paths, which can also vanish for $c_{XY}=-c_{XW} c_{WY}$, and seems like a more appropriate representation of the coupling between $X_{t-2}$ and $Y_t$.

Another example where one cannot unambiguously relate the coupling strength to one coefficient is for a nonlinear dependency between $X$ and $Y$ [Fig.~\ref{fig:sidepath_nonlinear}(b)]:
\begin{align}
Z_t &= \eta^Z_t  \nonumber \\
X_t &=  c_{ZX} Z_{t-1} + \eta^X_t  \nonumber \\
Y_t &= c_{XY} (X_{t-1})^2 + \eta^Y_t.
\end{align}
If we express $Y_t$ explicitly in terms of the source variance of $X$ and the parent of $X$
\begin{align}
Y_t &= c_{XY} c_{ZX}^2 (Z_{t-2})^2 + 2 c_{ZX} c_{XY} Z_{t-2} \eta^X_{t-1} +\nonumber\\
&~~~~~~~~~+ c_{XY} (\eta^X_{t-1})^2 + \eta^Y_t,
\end{align}
we note that due to the term $2 c_{ZX} c_{XY} Z_{t-2} \eta^X_{t-1}$ the effect of $Z_{t-2}$ is not additively separable from the source process $\eta^X_{t-1}$. In the Venn diagram of Fig.~\ref{fig:entropy_flow}(a) this ``mixing'' of entropies implies that the parts of the entropies $H(X|\mathcal{P}_X)$ and $H(\mathcal{P}_X)$ that overlap with $H(Y)$ are not distinguishable anymore, which could be visualized by the red and light gray shadings bleeding into one another. Therefore the coupling should be considered as emanating from $(X_{t-1},\mathcal{P}_{X_{t-1}})$ rather than $X_{t-1}$ alone [visualized by a thick arrow in Fig.~\ref{fig:sidepath_nonlinear}(b)].  For this nonlinear model we have not found an analytical expression for MIT, but the more general case of this model is studied numerically in the appendix.

These two examples point to constraints under which full coupling strength autonomy can be reached. In the next section we will formalize these constraints to general conditions in a theorem of coupling strength autonomy.

\section{Coupling strength autonomy theorem and modifications of MIT} \label{sect:coupling_independence}
Let $X$, $Y$ be two subprocesses of some multivariate stationary discrete-time process $\mathbf{X}$ sufficing condition~(S) in \cite{Eichler2011} with time series graph $G$ as defined in Sect.~\ref{sect:graph} and coupling link ``$X_{t-\tau}\to~Y_t$'' for $\tau>0$. The following derivations also hold for more than one link at lags $\tau'\neq\tau$ between $X$ and $Y$. As before, we denote their parents $\mathcal{P}_{Y_{t}}$ and  $\mathcal{P}_{X_{t}}$. For the link ``$X_{t-\tau}\to~Y_t$'' we define the following conditions:
\begin{enumerate}
\item \emph{Additivity} means that the dependence of $X_t$ on its source process $\eta^X_t$ and parents $\mathcal{P}_{X_{t}}$ and of $Y_t$ on its source process $\eta^Y_t$, $X_{t-\tau}$ and the remaining parents $\mathcal{P}_{Y_{t}}\setminus \{X_{t-\tau}\}$ is \emph{additive}, i.e., they can be written as
\begin{align}
X_t &=  g_X(\mathcal{P}_{X_{t}}) + \eta^X_t \label{eq:sepa_x} \\
Y_t &= f( X_{t-\tau}) + g_Y (\mathcal{P}_{Y_{t}}\setminus \{X_{t-\tau}\}) +  \eta^Y_t  \label{eq:sepa_y}
\end{align}
for possibly multivariate random variables $\mathcal{P}_{X_{t}}$ and $\mathcal{P}_{Y_{t}}\setminus \{X_{t-\tau}\}$, univariate i.i.d. random variables $\eta^X$ and $\eta^Y$ with arbitrary, not necessarily identical distributions, and arbitrary functions $g_Y,\,g_X,\,f$.

\item \emph{Linearity in f:} The dependence of $Y_t$ on $X_{t-\tau}$ is linear, i.e., $f(x)=c x$ with real $c$.

\item \emph{``No sidepath''-constraint}, i.e., in the time series graph $G$ the node $X_{t-\tau}$ is separated from $(\mathcal{P}_{Y_{t}} \setminus\mathcal{P}_{X_{t-\tau}})\setminus \{X_{t-\tau}\}$ given $\mathcal{P}_{X_{t-\tau}}$ (for a formal definition of paths and separation see \cite{Eichler2011}).
Since  due to condition~(S) in \cite{Eichler2011} separation implies conditional independence
\begin{align} \label{eq:nosidepath}
& I((\mathcal{P}_{Y_{t}} \setminus\mathcal{P}_{X_{t-\tau}})\setminus \{X_{t-\tau}\}; X_{t-\tau}|\mathcal{P}_{X_{t-\tau}})=0.
\end{align}
\end{enumerate}

\textbf{Theorem (Coupling Strength Autonomy).}
MIT defined in Eq.~(\ref{eq:def_mit}) for the coupling link ``$X_{t-\tau}\to~Y_t$'' for $\tau>0$ of a multivariate stationary discrete-time process $\mathbf{X}$ sufficing condition~(S) in \cite{Eichler2011} has the following dependency properties:
\begin{enumerate}

\item If all three conditions (1)-(3) hold, then MIT can be expressed as an MI of the source processes:
\begin{align} \label{eq:mit_1}
I^{\rm MIT}_{X{\to}Y}(\tau) = I(\eta^X_{t-\tau};c\eta^X_{t-\tau}+\eta^Y_t).
\end{align}
Since $\eta^Y_t$ and $\eta^X_{t-\tau}$ are assumed to be independent, the probability density of their sum is given by their convolution. The MIT thus depends solely on $c$ and the joint and marginal distributions of $\eta^X_{t-\tau}$ and the convolution of $c\eta^Y_t$ with $\eta^X_{t-\tau}$.

\item If only conditions (1) and (2) hold, i.e., there exists a sidepath between $X_{t-\tau}$ and some nodes in $\mathcal{P}_{Y_{t}} \setminus\mathcal{P}_{X_{t-\tau}}$, then MIT depends additionally on the distributions of at least the ``sidepath-parents'' in $\mathcal{P}_{Y_t}$ and their functional dependence on $Y_t$:
\begin{align} \label{eq:mit_2}
I^{\rm MIT}_{X{\to}Y}(\tau) = I(\eta^X_{t-\tau};c\eta^X_{t-\tau}+\eta^Y_t|\mathcal{P}_{Y_t}{\setminus}\{X_{t-\tau}\}).
\end{align}
This relation can be further simplified if $g_Y (\mathcal{P}_{Y_{t}}\setminus \{X_{t-\tau}\})$ is additive in some parents.

\item If only the additivity condition (1) holds, i.e., $f(x)$ is nonlinear and mixes $\eta^X_{t-\tau}$ with the parents $\mathcal{P}_{X_{t-\tau}}$
then MIT depends additionally on $f$, the distributions of variables in $\mathcal{P}_{X_{t-\tau}}$ as well as $\mathcal{P}_{Y_t}{\setminus}\{X_{t-\tau}\}$ and their functional dependencies on $Y_t$:
\begin{align} \label{eq:mit_3}
&I^{\rm MIT}_{X{\to}Y}(\tau) = \nonumber \\
&=I(\eta^X_{t-\tau};\,f(\eta^X_{t-\tau}+g_X(\mathcal{P}_{X_{t-\tau}}))+\eta^Y_t~|\nonumber\\
&~~~~~~~~~~~~~~~~~~~~~~~~~~~~~|~\mathcal{P}_{Y_t}{\setminus}\{X_{t-\tau}\},\,\mathcal{P}_{X_{t-\tau}}).
\end{align}
This relation can be further simplified if some parents in $\mathcal{P}_{Y_t}{\setminus}\{X_{t-\tau}\}$ are independent of $f(\eta^X_{t-\tau}+g_X(\mathcal{P}_{X_{t-\tau}}))$.
\end{enumerate}
For a contemporaneous link ``$X_t\--~Y_t$'' the contemporaneous MIT defined in Eq.~(\ref{eq:def_mit_contemp}) under the condition (1) is:
\begin{align} \label{eq:mit_contemp}
I^{\rm MIT}_{X\--Y} = I(\eta^X_{t};\eta^Y_t|\mathcal{N}_{X_t}{\setminus}\{Y_t\},\mathcal{N}_{Y_t}{\setminus}\{X_t\}).
\end{align}
A contemporaneous link cannot have sidepaths. For $X=Y$ MIT measures the autodependency strength.
The proofs are given in the appendix.

We now discuss some remarks on the theorem and possible modifications of MIT: 
\begin{enumerate}[i)]
\item For the special case of multivariate linear autoregressive processes of order $p$  \cite{brockwell2009time} defined by 
\begin{align} \label{eq:var}
\mathbf{X}_t &= \sum_{s=1}^p \mathbf{\Phi}(s) \mathbf{X}_{t-s} + \mathbf{\varepsilon}_t, ~~~~\mathbf{\varepsilon}_t\sim \mathcal{N}(0,\Sigma),
\end{align}
with the coupling coefficient $c_{XY}$ at lag $\tau$ corresponding to the connectivity matrix entry $\mathbf{\Phi}(\tau)_{YX}$, and with no sidepaths, Eq.~(\ref{eq:mit_1}) leads to
\begin{align} \label{eq:mit_var}
I^{\rm MIT}_{X{\to}Y}(\tau) &= \frac{1}{2} \ln \left( 1 + \frac{c_{XY}^2\sigma^2_X}{\sigma_Y^2} \right),
\end{align}
generalizing the MIT for our analytical model in Eq.~(\ref{eq:mit}). For an autodependency at lag $\tau$ with coefficient $a_Y$ and no sidepaths the MIT is $I^{\rm MIT}_{Y{\to}Y}(\tau) = \frac{1}{2} \ln \left( 1 + a_{Y}^2 \right)$, independent of the source variance $\sigma^2_Y$.

\item
The form Eq.~(\ref{eq:mit_var}) is reminiscent of the Shannon-Hartley theorem in communication theory \cite{Cover2006}. There the coupling strength corresponds to the communication channel capacity $C$ which is given by the maximum MI over all possible input sources: $C = \max_{\{ P(X) \}} I(X;Y)$. The Shannon-Hartley theorem for Gaussian channels then reads
\begin{align}
C = B \log \left( 1 + \frac{S}{N} \right)
\end{align} 
with bandwidth $B$ and signal-to-noise ratio $S/N$, which in Eq.~(\ref{eq:mit_var}) corresponds to $c_{XY}^2\sigma^2_X/\sigma_Y^2$. The difference to our measure of coupling strength is that we cannot manipulate the input sources and thus cannot measure the channel capacity alone. We also expressed the various other CMIs occuring above in this form, where the quotient can be interpreted as a signal-to-noise ratio. For example, in Eq.~(\ref{eq:px}) $c_{XY}^2\sigma_X^2$ is the signal strength and $c_{WY}^2 \sigma_W^2+\sigma_Y^2$ is the noise strength.

\item
For sidepaths, i.e., under the conditions (1) and (2) only, the example of MIT and the modified MIT for the case of our model example Eq.~(\ref{eq:ana_model_sidepath}) point to the suggestion, that it might be more appropriate to ``leave open'' all paths from $X_{t-\tau}$ to $Y_t$ by excluding those parents of $Y_t$ that are depending on $X_{t-\tau}$, i.e., 
\begin{align}
\mathcal{P}^\star_{Y_t}\equiv \{W^k_{t-\tau_k} \in \mathcal{P}_{Y_t}{\setminus}\mathcal{P}_{X_{t-\tau}}: I(W^k_{t-\tau_k}; X_{t-\tau}|\mathcal{P}_{X_{t-\tau}})>0 \},
\end{align}
but additionally including the parents $\mathcal{P}(\mathcal{P}^\star_{Y_t})$ of these sidepath parents. In this way the couplings via the direct link ``$X_{t-\tau}\to~Y_t$'' and the path  ``$X_{t-\tau}\mathop{}_{\--}^{\to}\mathcal{P}^\star_{Y_t}\to Y_t$'' (the symbol ``$\mathop{}_{\--}^{\to}$'' denotes that the link from $X_{t-\tau}$ to the sidepath parents can either be directed or contemporaneous) are isolated from the effects of their parents. The modified MIT we call MITS where ``S'' stands for ``sidepath'':
\begin{align} \label{eq:mit_star}
I^{\rm MITS}_{X{\to}Y}(\tau)&\equiv I(X_{t-\tau};Y_t~|~ \mathcal{P}_{Y_t}{\setminus} \{ \mathcal{P}^\star_{Y_t},X_{t-\tau}\},\nonumber\\
                            &~~~~~~~~~~~~~~~~~~~~~~~~~~ \mathcal{P}(\mathcal{P}^\star_{Y_t}) {\setminus}\{X_{t-\tau} \}, \mathcal{P}_{X_{t-\tau}} ).
\end{align}

\item
For nonlinear dependencies $f$ one could modify MIT to the CMI between $Y_t$ and the joint vector $(X_{t-\tau},\,\mathcal{P}_{X_{t-\tau}})$ leading to MITN where ``N'' stands for ``nonlinear'':
\begin{align} \label{eq:mit_starstar}
I^{\rm MITN}_{X{\to}Y}(\tau)&\equiv I((X_{t-\tau},\,\mathcal{P}_{X_{t-\tau}});Y_t~|~ \mathcal{P}_{Y_t} {\setminus}(X_{t-\tau},\mathcal{P}_{X_{t-\tau}} )).
\end{align}
\end{enumerate}
These modifications will be studied in a separate paper.

The theorem implies that under the conditions (1)-(3) the MIT is independent of other coefficients belonging to other links. If this holds for all coupling strengths of all links in the model, then the MITs are independent in a functional sense. Note, however, that all coupling strengths of links emanating from the same process $X$ will depend on the source variance of $\eta^{X}$.
Thus, MIT somewhat disentangles the coupling structure, which is exactly the coupling strength autonomy that makes MIT well interpretable as a measure that solely depends on the ``coupling mechanism'' between $X$ at lag $t-\tau$ and $Y_t$, autonomous of other processes.  
One such possible misleading input ``filtered out'' by MIT is autocorrelation, or, more generally, autodependency as will be shown in the numerical experiments and the application to climatological data. 
In the next section we investigate the coupling strength autonomy property numerically.

\section{Numerical Comparison} \label{sect:numerics}
In the following we compare MI, TE, MIT and related measures numerically to investigate the properties of generality and coupling strength autonomy for a general class of nonlinear discrete-time stochastic multivariate processes:
\begin{align} \label{eq:model}
   Z_t &= a_Z Z_{t-1} +  \eta^{Z}_t \nonumber \\
   X_t &= a_X X_{t-1} + c_{ZX}~g(Z_{t-1}) +  \eta^{X}_t \nonumber  \\
   Y_t &= a_Y Y_{t-1} + c_{WY}~g(W_{t-1}) + c_{XY}~f(X_{t-2}) +  \eta^{Y}_t \nonumber\\
   W_t &= a_W W_{t-1} +  \eta^{W}_t
\end{align}
with independent Gaussian white noise processes $\eta^{\cdot}_t$ with all variances $\sigma^2_{\cdot}=1$. The corresponding time series graph is depicted in Fig.~\ref{fig:entropy_flow}(b). We estimate the various coupling measures for fixed $c_{XY}$ and $a_Z=a_W=0.5$ and vary the input coefficients
\begin{align*}
a_X=c_{ZX} \in \{0.0,\,0.1,\,\ldots,\,0.8\} \\
a_Y=c_{WY} \in  \{0.0,\,0.1,\,\ldots,\,0.8\} 
\end{align*}
and functional dependencies of inputs
\begin{align*} 
\text{linear}~~~g(x)&=x,\\
\text{squared}~~~g(x)&= 0.3 \cdot x^2,\\
\text{stochastic}~~~g(x)&=2x\varepsilon_t~~~\text{with uniform i.i.d. $\varepsilon_t \in [0,1]$},\\
\text{exponential}~~~g(x)&=0.3 \cdot 2^x,\\
\text{sinusoidal}~~~g(x)&=\sin 4x.
\end{align*}
Here we depict results for linear $f(x)=x$ such that the multivariate process suffices all three conditions, a nonlinear dependency type is discussed in the appendix. The ensemble $E$ then consists of all combinations of input coefficients and functional forms, each combination run with 120 trials. The CMIs are estimated using a nearest-neighbor ($k$NN) estimator \cite{Kraskov2004a,FrenzelPompe2007} with parameter $k=1$ (small values of $k$ lead to a lower estimation bias but higher variance \cite{Kraskov2004a,FrenzelPompe2007}). 
\begin{figure}[t]
\includegraphics[width=\columnwidth]{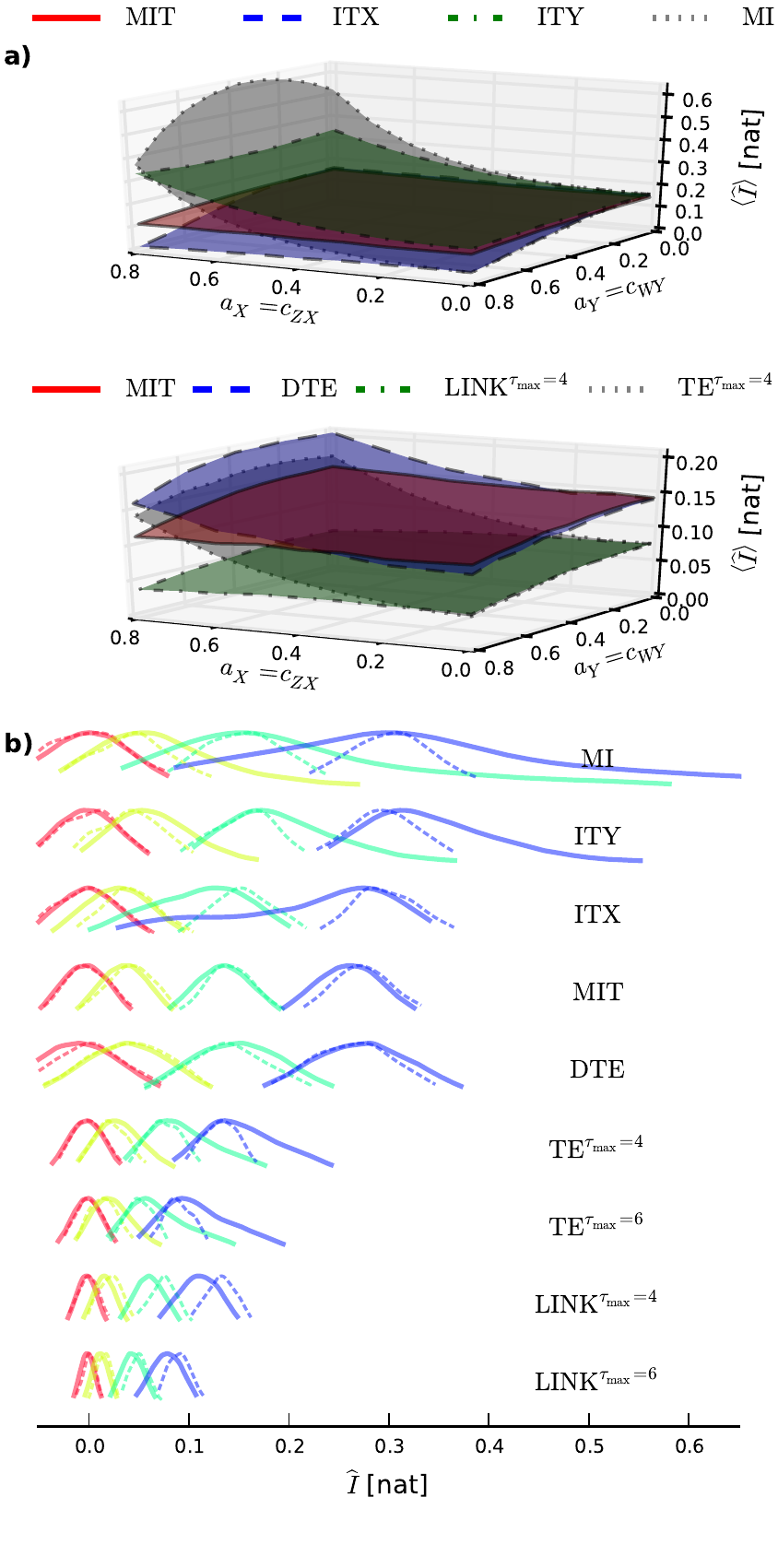}
\caption[]{(Color online) Numerical experiments with the model Eq.~(\ref{eq:model}) using time series length $T=1000$. In (a) we plot the ensemble average $\left< I \right>_E$ for fixed $c_{XY}=0.6$ for all measures as specified in the main text. 
In (b) we show the ensemble densities of all measures for different coupling coefficients $c_{XY}=0.0,\,0.3,\,0.6,\,0.9$ (from left to right red, yellow, green and blue solid lines).
The densities are estimated using Gaussian kernel smoothing according to Scott's rule, showing only the 90\% most probable ensemble members.
}
\label{fig:model}
\end{figure}

In the top panel of Fig.~\ref{fig:model}(a) we plot the ensemble average $\left< \hat{I} \right>_E$ for fixed $c_{XY}=0.6$ for the following measures with $\tau=2$: MI $I(X_{t-\tau};Y_t)$ (gray with dotted line), ITY according to Eq.~(\ref{eq:def_py}) (green with dash-dotted line), ITX according to Eq.~(\ref{eq:def_px})  (blue with dashed line) and MIT according to Eq.~(\ref{eq:def_mit}) (red with solid line). The parents are shown in Fig.~\ref{fig:entropy_flow}(b). 

MIT is largely invariant to changes of the remaining coefficients and $g(x)$ and approximately attains the analytical value for zero input coefficients [given by Eq.~(\ref{eq:mit}) for $c_{XY}=0.6$ and $\sigma^2_X=\sigma^2_Y=1$]: $I\approx0.15$. This implies that the MIT of the coupling link is autonomous of the MITs corresponding to the input links $Z{\to}X$ for $Z\in \mathcal{P}_X$ and $W{\to}Y$ for $W\in \mathcal{P}_Y{\setminus}\{X\}$  which scale with these coefficients. Note, however, that all coupling strengths of links emanating from the same process will depend on its variance $\sigma^2_{\cdot}$ like in Eq.~(\ref{eq:mit}). Further, MI is mostly larger, but can also be smaller than MIT, which can be explained with the entropy diagram in Fig.~\ref{fig:entropy_flow}(a): larger MIs occur if the entropy is increased due to a larger input of $H(\mathcal{P}_{X})$ and smaller MIs occur if the relative shared part of $H(X)$ in $H(Y)$ decreases due to a larger input of $H(\mathcal{P}_Y)$. For zero inputs, MI approaches the analytical value $I\approx0.15$ where all four measures converge to. ITY can at least exclude input to $Y$ and ITX can exclude input to $X$. Note, however, that the dependence of ITX and ITY on the input coefficients can be different in other models. The average of ITX (ITY) is always smaller (larger) equal than MIT confirming the inequality Eq.~(\ref{eq:inequality}). 

In the bottom panel of Fig.~\ref{fig:model}(a) we compare MIT (red with solid line) to TE according to Eq.~(\ref{eq:te_trunc}) truncated at $\tau_{\max}=4$ (gray with dotted line), the CMI $I^{\rm LINK}_{X\to Y}$ defining links  in the time series graph according to Eq.~(\ref{eq:def_graph}) truncated at $\tau_{\max}=4$ (green with dash-dotted line), and DTE according to Eq.~(\ref{eq:sum}) with $\tau^\star=3$ (blue with dashed line).
TE and LINK have a much larger estimation dimension of 17 (as much as 25 for $\tau_{\max}=6$) compared to 6 for MIT and between 5 and 12 for the summands of DTE. Compared to DTE this leads to a negative relative bias in TE of about 50\% for the analytically known value for zero input coefficients $I\approx 0.15$.  Apart from this bias, TE and DTE scale similarly with the input coefficients. LINK is dependent on $a_X$ as we expect from our analytical considerations [Eq.~(\ref{eq:link_anamodel})]. The MIT shows some slight dependence for strong inputs due to estimation problems for short samples, but otherwise also numerically we demonstrate here that only MIT fulfills the proposed property of coupling strength autonomy.

In Fig.~\ref{fig:model}(b) we show the whole densities of $E$ of all measures for \textit{different} coupling coefficients $c_{XY}$. The aim of this experiment is to measure how well the measures can distinguish the coupling strength for different $c_{XY}$ as demanded by the property of equitability. The dashed lines show the densities of the ensemble for $a_X=c_{ZX}=a_Y=c_{WY}=0$, i.e., if both $X$ and $Y$ are independent of their parents.
   
As we now already expect, MI takes a whole range of values for the same $c_{XY}$. ITY is broadly  peaked towards higher $I$ values and ITX towards lower values, confirming the inequality Eq.~(\ref{eq:inequality}). Note, that this relation holds only on average. Only with MIT the different coupling coefficients $c_{XY}$ can be well distinguished. DTE tends to slightly higher values for larger autodependencies within $X$ as expected from our analytical results. Additionally, the variance of the DTE estimate is higher because each summand's variance adds up to the total variance of the DTE estimate. The remaining four plots demonstrate that TE and the CMI of Eq.~(\ref{eq:def_graph}) strongly suffer from the negative bias associated with high dimensional estimation depending on the chosen $\tau_{\max}$. TEs or LINKs estimated with different $\tau_{\max}$ can, therefore, not be compared with each other. 

For the `unperturbed' case of zero inputs, the ensemble distributions of MI [dashed lines in Fig.~\ref{fig:model}(b)] are -- as expected -- similar to the one for MIT with ``conditioned-out'' inputs (solid lines) apart from a small bias and smaller variance related to slightly higher dimensional estimation. 
For conditionally independent variables ($c_{XY}=0$, red lines), all measures have almost no bias, i.e., $\hat{I}\approx 0$, which is a property of the $k$NN estimator and holds also for short samples \cite{Kraskov2004a}. It may seem that apart from the bias, at least the variance is much smaller for the high dimensional measures TE and LINK, but the relative variance $\left<\hat{I}^2\right>/\left<\hat{I}\right>$ actually increases leading to a worsened distinguishability.

Summarizing, our experiments provide numerical evidence that MIT acts as an information-theoretic ``filter'' that excludes undesired effects of autodependency or other misleading inputs. The MIT is, thus, specific only to the interaction of the two lagged subprocesses and can disentangle the measured coupling strengths of the different links in a time series graph. The commonly used measures MI and TE, on the other hand, are possibly affected also by the interactions that $X$ and $Y$ have with other processes. In this respect MIT is more intuitive and better interpretable than TE or MI. The coupling strength autonomy property can, thus, be regarded as one ingredient of a multivariate extension of the equitability property.

\section{Discussion and Limitations} \label{sect:limitations}
Let us here discuss some limitations of our approach:
\begin{enumerate}[i)]
\item Our notion of causality is to be understood only with respect to the observable processes included in the parents, while the general notion of causality \cite{Pearl2000} requires to exclude the influence of the whole universe.

\item The graphical model imposes a discrete description of causal interactions. 
Regarding the source entropy, we face the problem that if a time-continuous process is sampled at some interval $\Delta s$, there is an infinite set of unobserved nodes in between every $X_{t}$ and $X_{t-1}$ for $X\in \mathbf{X}$ in the time series graph. We will, therefore, not be able to access the source entropy solely at time $t$, but only the aggregated information in the interval $[t-\Delta s,t]$. But for discrete processes graphical models are applicable to the large class of models sufficing condition~(S) in \cite{Eichler2011}.

\item Although the graphical model approach reduces the estimation dimension to a minimum, the dimension can still be relatively high leading to biased estimates for shorter samples. A study on the effects of high dimensional estimation is subject to further research.
Generally, there are problems with entropy estimation for highly skewed distributions which need to be resolved by improved estimators of CMI.

\item Our two-step approach first necessitates the estimation of the time series graph which comes with the associated problems of false positive detections due to multiple testing and missed causal links. These problems are analyzed in the Supplementary Material in \cite{Runge2012prl}.

\item As discussed in the coupling strength autonomy theorem, not in all cases a coupling strength can be attributed to only one single coefficient. Only if this is the case, i.e., under the conditions (1)-(3), MIT can filter out all influences from the parents of $X$ and $Y$. If the dependency is nonlinear or sidepaths exist, one could use modifications of MIT like $I^{\rm MITS}_{X{\to}Y}$ [Eq.~(\ref{eq:mit_star})] and $I^{\rm MITN}_{X{\to}Y}$ [Eq.~(\ref{eq:mit_starstar})] for a more appropriate measure of coupling strength. Note, that even so for full coupling strength autonomy the link ``$X_{t-\tau}\to~Y_t$'' needs to be linear, the remaining dependencies can still be nonlinear and the source processes can have arbitrary distributions. The process can, therefore, not easily be estimated using model-based regressions. 

\item Regarding equitability, a desired property of a coupling measure would be that it scales linearly with the coupling parameter $c_{XY}$ like the partial correlation approximately in the Gaussian case. As can be seen from the analytical derivations and the numerical example in Fig.~\ref{fig:model}(b), MIT scales $\propto \ln (1+c_{XY}\cdots)$ for Gaussian dependencies, but a linear scaling in this case can be attained by the transformation $I \to\sqrt{1-e^{-2I}}$\cite{Cover2006}. For more complex dependencies improved estimators that are more adapted to the distributions might help.
\end{enumerate}

\section{Application to climatological time series} \label{sect:application}
\begin{figure}[t]
\begin{center}
\includegraphics[width=\columnwidth]{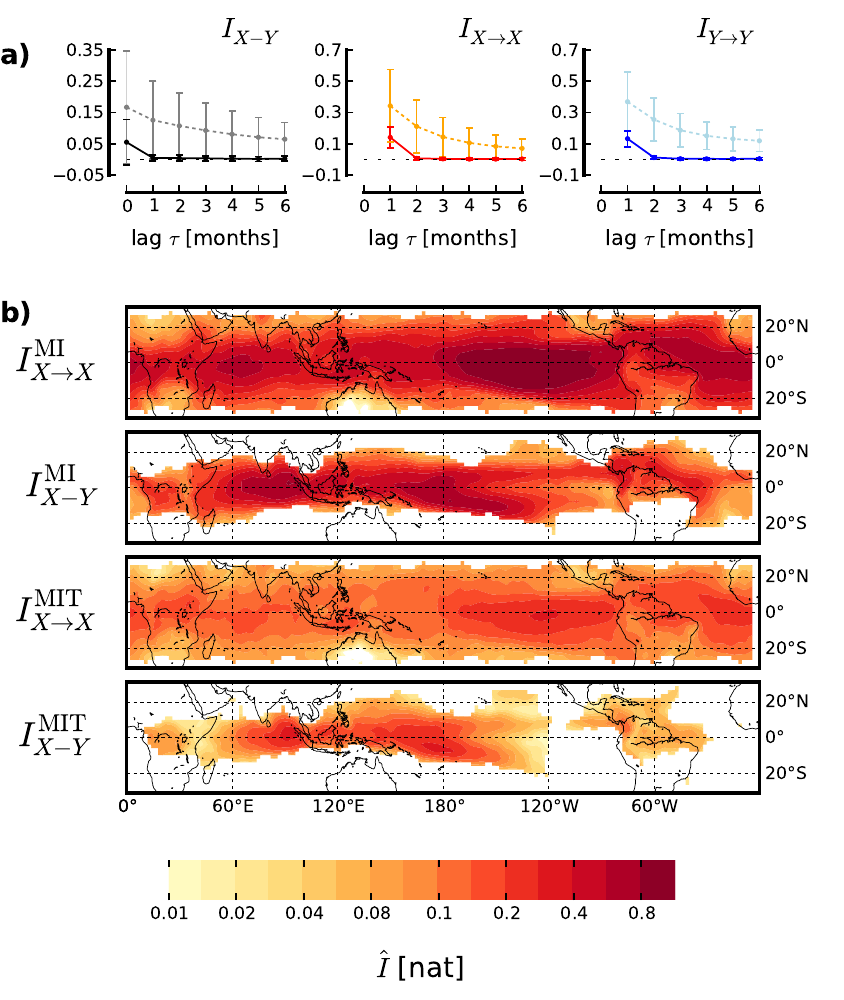}
\end{center}
\caption{(Color online) Analysis of air temperature anomalies at the surface ($X$) and the upper troposphere ($Y$), $T= 1008$ months (1927-2011). (a) shows the spatial average and standard deviation of coupling (left plot) and autodependency (middle plot for $X$, right plot for $Y$) lag functions for MI (dashed lines in light colors) and MIT (solid lines in dark colors). In (b) we spatially resolve the coupling strengths of the contemporaneous link ``$X_t\--~Y_t$'' and the autodependency ``$X_{t-1}\to X_t$'' for MI (upper two panels) and MIT (lower two panels). $I^{\rm MI}_{Y{\to}Y}$ and $I^{\rm MIT}_{Y{\to}Y}$ (not shown) are almost the same all across the tropics. For the contemporaneous link values below the $98\%$ significance level are in white. CMIs estimated with $k=10$.
}
\label{fig:application}
\end{figure}

We now analyze monthly air temperature anomalies in the tropics at two different altitudes in a NCEP/NCAR reanalysis data set \cite{Compo2006}. To investigate the upwelling of heat from the sea surface towards the upper troposphere in a height of about 12 km, we measure the coupling strength between the surface pressure level ($X$ in Fig.~\ref{fig:application}) and the 200 hPa pressure level ($Y$) for all tropical (latitudes between 30$^{\rm o}$S and 30$^{\rm o}$N) grid points. 

First, we estimated the time series graph using the algorithm introduced in \cite{Runge2012prl} separately for each surface-troposphere pair at each grid point using a significance threshold estimated with the shuffle test as in \cite{Runge2012prl}. We found -- on average -- the parents $\mathcal{P}_{X_t}=\{X_{t-1}\}$ and $\mathcal{P}_{Y_t}=\{Y_{t-1}\}$, i.e., lag-1 autodependencies, and the contemporaneous link ``$X_t\--~Y_t$''.
 
With these parents, the spatial average of all lag functions of MIT in the left panel of Fig.~\ref{fig:application}(a) shows the contemporaneous link ``$X_{t}\--~Y_t$'' as a significant peak, indicating that the time scale of the coupling is below the lag of one month. The MI, on the other hand, is significant for a wide range of lags, making an assessment of a physical coupling delay difficult. While the contemporaneous link cannot be interpreted as a directed coupling, we can still assess its strength. The MIT of a linear Gaussian process with the same time series graph is $I^{\rm MIT}_{X{\--}Y} = \frac{1}{2} \log \left( \frac{\sigma^2_X \sigma^2_Y}{\sigma^2_X \sigma^2_Y-\sigma_{XY}^2} \right)$, while MI additionally depends on the autodependency coefficients.

Figure~\ref{fig:application}(b) shows a large (compared to the extra tropics) $I^{\rm MI}_{X{\--}Y}$ all across the tropics. Significant $I^{\rm MIT}_{X{\--}Y}$ values, on the other hand, are more confined and largest between 90$^{\rm o}$E and 170$^{\rm o}$W. Larger MIT values indicate a stronger coupling between the surface and upper tropospheric level in an area that actually corresponds to a region of strong upwelling in the Walker circulation \cite{Lau2002}. The difference between MI and MIT is largest in the Eastern Pacific where also the increased autodependency in surface air temperatures is apparent ($I^{\rm MIT}_{X{\to}X}$). This strong persistence thus leads to a spurious increase in MI, which cannot differentiate the effects of increased autodependencies and increased contemporaneous coupling like MIT. With our measure of coupling strength we are, thus, able to infer a more reasonable picture of the physical interactions in the Walker circulation. This preliminary example underlines the importance of having a meaningfully interpretable coupling measure.

\section{Conclusions}
To conclude, we have analytically and numerically shown that the commonly used measures MI and TE can be rather unintuitive as measures of coupling strength. To overcome this limitation, we propose a two-step approach, where in the first step the existence of lag-specific couplings, i.e., the causal links, and contemporaneous links in a multivariate process are determined as discussed in \cite{Runge2012prl}. For the second step addressed in the present article, we have generalized the information-theoretic MIT as a lag-specific measure that has a property which we call coupling strength autonomy. It allows for a well interpretable coupling strength reminiscent of an experimentally manipulable setting. As we prove analytically and numerically, the coupling strength autonomy property is useful for models of processes where the coupling strength can be attributed to one single coefficient, while for other cases we suggest modifications of MIT as more appropriate measures. Compared to TE, our MIT has the advantage of being practically computable without the need for arbitrary truncations.
Besides our example from climatology, also in other fields of science our two-step approach promises to not only extract the causal direct (rather than the indirect) connectivity among processes, but also to assess a meaningful coupling strength, that -- together with the coupling delay -- assists a physical interpretation.

\section*{Acknowledgments}
We appreciate the support by the German National Academic Foundation, the DFG grant No. KU34-1, the DFG research group 1380 ``HIMPAC'', and the German Federal Ministry for Education and Research (BMBF) via the Potsdam Research Cluster for Georisk Analysis, Environmental Change and Sustainability (PROGRESS). We thank Lara Neureither for helpful comments.

\section*{Appendix}
Here we give the proofs of the inequality relation between MIT, ITX and ITY in Eq.~(\ref{eq:inequality}), the coupling strength autonomy theorem and further discussions regarding the property of coupling strength autonomy for processes violating the linearity condition (2).

\renewcommand{\theequation}{A\arabic{equation}}
\setcounter{equation}{0}
\setcounter{section}{0}
\section{Proof of inequality relation Eq.~(\ref{eq:inequality})}
\noindent
The MIT $I^{\rm MIT}_{X{\to}Y}=I(X_{t-\tau};Y_t|\mathcal{P}_{Y_t}{\setminus}\{X_{t-\tau}\},\mathcal{P}_{X_{t-\tau}})$ between two uni- or multivariate subcomponents $X,Y$ of a stationary multivariate discrete-time stochastic process $\mathbf{X}$ with time series graph $G$ and  parents $\mathcal{P}$ as defined in the main article, is bounded by the two CMIs with condition on either parents [Eq.~(\ref{eq:inequality})]
\begin{align} 
 I(X_{t-\tau};Y_t|\mathcal{P}_{X_{t-\tau}}) \leq I^{\rm MIT}_{X{\to}Y} \leq  I(X_{t-\tau};Y_t|\mathcal{P}_{Y_t}{\setminus}\{X_{t-\tau}\}).
\end{align}
where $\tau>0$. The right inequality holds for all processes sufficing the very general condition\,(S) in \cite{Eichler2011} and the left inequality if additionally the ``no sidepath''-constraint for the coupling ``$X_{t-\tau} \to Y_t$'' holds, that is, if $X_{t-\tau}$ is separated from $\mathcal{P}_{X_{t-\tau}} \setminus \mathcal{P}_{Y_t}$ by its parents $\mathcal{P}_{X_{t-\tau}}$ in the time series graph. For a definition of separation see \cite{Eichler2011}.

To prove the right inequality, let $\tilde{\mathcal{P}}_{X_{t-\tau}}$ be the set of parents of $X_{t-\tau}$ that is not already included in $\mathcal{P}_{Y_t}$, i.e., $\tilde{\mathcal{P}}_{X_{t-\tau}}=\mathcal{P}_{X_{t-\tau}} \setminus\mathcal{P}_{Y_t}$. Then it holds that $I(\tilde{\mathcal{P}}_{X_{t-\tau}}; Y_t|\mathcal{P}_{Y_t})=0$ because the parents $\mathcal{P}_{Y_t}$ separate $Y_t$ from any subset of $\mathbf{X}_t^-\setminus \mathcal{P}_{Y_t}$ and separation in the time series graph implies conditional independence between the subprocesses \cite[Thm.\,4.1]{Eichler2011}. Now we apply the chain rule on the (multivariate) CMI $I(X_{t-\tau}, \tilde{\mathcal{P}}_{X_{t-\tau}};Y_t| \mathcal{P}_{Y_t} \setminus \{X_{t-\tau}\})$ twice:
\begin{align*}
& I(X_{t-\tau},\tilde{\mathcal{P}}_{X_{t-\tau}};Y_t| \mathcal{P}_{Y_t} \setminus \{X_{t-\tau}\}) = \\ 
&= I(X_{t-\tau};Y_t| \mathcal{P}_{Y_t} \setminus \{X_{t-\tau}\}) + 
\underbrace{I(\tilde{\mathcal{P}}_{X_{t-\tau}};Y_t| \mathcal{P}_{Y_t} )}_{=0} \\
& = \underbrace{I(\tilde{\mathcal{P}}_{X_{t-\tau}};Y_t| \mathcal{P}_{Y_t} \setminus \{X_{t-\tau}\})}_{\geq 0}+\\
   & ~~~~+I(X_{t-\tau};Y_t| \mathcal{P}_{Y_t} \setminus \{X_{t-\tau}\},\tilde{\mathcal{P}}_{X_{t-\tau}}) \\
& \implies  I(X_{t-\tau};Y_t|\mathcal{P}_{Y_t}{\setminus}\{X_{t-\tau}\},\mathcal{P}_{X_{t-\tau}}) \\
            &~~~~~~~~~~~\leq I(X_{t-\tau};Y_t|\mathcal{P}_{Y_t}{\setminus}\{X_{t-\tau}\}).
\end{align*}
Note, that (conditional) mutual information is always non-negative.

For the left inequality we now define $\tilde{\mathcal{P}}_{Y_{t}}$ to be the set of parents of $Y_{t}$ that is not already included in $\mathcal{P}_{X_{t-\tau}}$, i.e., $\tilde{\mathcal{P}}_{Y_{t}}=\mathcal{P}_{Y_{t}} \setminus\mathcal{P}_{X_{t-\tau}}$. 
Then under the ``no sidepath''-constraint it holds that $I(\tilde{\mathcal{P}}_{Y_{t}}\setminus \{X_{t-\tau}\}; X_{t-\tau}|\mathcal{P}_{X_{t-\tau}})=0$.
Note, that all paths emanating from $X_{t-\tau}$ towards the past are surely blocked by $\mathcal{P}_{X_{t-\tau}}$ because they contain the motifs ``$\to Z_{t-\tau'}\to X_{t-\tau}$'' or ``$\-- Z_{t-\tau'}\to X_{t-\tau}$'' which are both blocked as $Z_{t-\tau'}\in \mathcal{P}_{X_{t-\tau}}$. The ``no sidepath''-constraint further demands that there are no unblocked paths to $\tilde{\mathcal{P}}_{Y_{t}}$ emanating towards the present or future. 
Again, we apply the chain rule on the (multivariate) CMI $I(X_{t-\tau};Y_t,\tilde{\mathcal{P}}_{Y_{t}}\setminus \{X_{t-\tau}\}| \mathcal{P}_{X_{t-\tau}})$ twice:
\begin{align*}
& I(X_{t-\tau};Y_t,\tilde{\mathcal{P}}_{Y_{t}}\setminus \{X_{t-\tau}\}| \mathcal{P}_{X_{t-\tau}}) = \\ 
&= I(X_{t-\tau};Y_t|  \mathcal{P}_{X_{t-\tau}}) + 
\underbrace{ I(X_{t-\tau};\tilde{\mathcal{P}}_{Y_{t}}\setminus \{X_{t-\tau}\}| \mathcal{P}_{X_{t-\tau}}, Y_t) }_{\geq 0} \\
& = \underbrace{I(\tilde{\mathcal{P}}_{Y_{t}}\setminus \{X_{t-\tau}\}; X_{t-\tau}|\mathcal{P}_{X_{t-\tau}})}_{= 0}+\\
   & ~~~~+I(X_{t-\tau};Y_t| \tilde{\mathcal{P}}_{Y_t} \setminus \{X_{t-\tau}\},\mathcal{P}_{X_{t-\tau}}) \\
& \implies  I(X_{t-\tau};Y_t|\mathcal{P}_{Y_t}{\setminus}\{X_{t-\tau}\},\mathcal{P}_{X_{t-\tau}}) \\
            &~~~~~~~~~~~\geq I(X_{t-\tau};Y_t|\mathcal{P}_{X_{t-\tau}}).
\end{align*}

\section{Derivations for analytical model Eq.~(\ref{eq:ana_model})}
Defining variances and covariances by
\begin{align} 
\Gamma_{ij}(\tau) \equiv E[\mathbf{X}^i_{t+\tau}  \mathbf{X}^j_t],
\end{align}
for model Eq.~(\ref{eq:ana_model}) the variances are 
\begin{align*}
\Gamma_X &= \frac{\sigma _X^2}{1{-}a^2_X} \\
\Gamma_Z &= c_{XZ}^2 \Gamma_X + \sigma_Z^2 \\
\Gamma_W &= \sigma_W^2 \\
\Gamma_Y &= c_{XY}^2 \Gamma_X + c_{WY}^2 \Gamma_W + \sigma_Y^2.
\end{align*}
Further, auto-covariances are
\begin{align*}
\Gamma_{XX}(\tau) &= a_X^{|\tau|} \Gamma_X \\
\Gamma_{YY}(\tau) &= c_{XY} \Gamma_{XX}(\tau) \\
\Gamma_{ZZ}(\tau) &= c_{XZ} \Gamma_{XX}(\tau) \\
\Gamma_{WW}(\tau) &= 0,
\end{align*}
with $\Gamma_{XX}(\tau=0) \equiv \Gamma_X$. The covariances for $\tau \geq 0$ are given by
\begin{align*}
\Gamma_{YX}(\tau) &= c_{XY} \Gamma_{XX}(\tau-2) \\
\Gamma_{XY}(\tau) &= a_X c_{XY} \Gamma_{XX}(\tau+1) \\
\Gamma_{ZX}(\tau) &= c_{XZ} \Gamma_{XX}(\tau-1) \\
\Gamma_{XZ}(\tau) &= a_X c_{XZ} \Gamma_{XX}(\tau) \\
\Gamma_{XW}(\tau) &= \Gamma_{WX}(\tau)=0 \\
\Gamma_{ZY}(\tau) &= c_{XY} c_{XZ} \Gamma_{XX}(\tau+1) \\
\Gamma_{YZ}(\tau) &= c_{XY} c_{XZ} \Gamma_{XX}(\tau-1) \\
\Gamma_{ZW}(\tau) &= \Gamma_{WZ}(\tau)=0 \\
\Gamma_{YW}(\tau) &= c_{WY} \delta(\tau-1) \Gamma_{W},  \\
\Gamma_{YW}(\tau) &= 0,
\end{align*}
with the Kronecker-Delta $\delta(s)=1$ for $s=0$ and $\delta=0$ else. These covariances form the entries of the covariance matrices that are needed to compute the conditional entropies.

\subsection{Derivations of TE}
For the derivation of TE
\begin{align*}
I^{\rm TE}_{X\to Y} &= H(Y_t|Y^-_t,W^-_t,Z^-_t) - H(Y_t|X^-_t Y^-_t,W^-_t,Z^-_t)
\end{align*}
we know from Markov properties that the latter term is the source entropy $H(Y_t|\mathcal{P}_{Y_t})=\frac{1}{2} \ln 2 \pi e \sigma^2_Y$. For the first entropy 
\begin{align} \label{eq:ana_model_ent}
H(Y_t|Y^-_t,W^-_t,Z^-_t) &= \frac{1}{2} \ln \left( 2 \pi e \frac{|\Gamma_{Y_t Y^-_t W^-_t Z^-_t}|}{|\Gamma_{Y^-_t W^-_t Z^-_t}|} \right)
\end{align}
we can write the covariance as a block matrix
\begin{align}
\Gamma_{Y_t Y^-_t W^-_t Z^-_t} &= 
\left(\begin{array}{cccc}
\Gamma_{Y_t}            & \Gamma_{Y_t; Y^-_t}        & \Gamma_{Y_t; W^-_t}         & \Gamma_{Y_t; Z^-_t} \\
\Gamma_{Y_t; Y^-_t}^\top & \Gamma_{Y^-_t }           & \Gamma_{Y^-_t; W^-_t}       & \Gamma_{Y^-_t; Z^-_t} \\
\Gamma_{Y_t; W^-_t}^\top & \Gamma_{Y^-_t; W^-_t}^\top & \Gamma_{W^-_t}              & \Gamma_{W^-_t; Z^-_t} \\
\Gamma_{Y_t; Z^-_t}^\top & \Gamma_{Y^-_t; Z^-_t}^\top & \Gamma_{W^-_t; Z_t^-}^\top      & \Gamma_{ Z^-_t} 
\end{array}
\right).
\end{align}
where, e.g., $\Gamma_{Y_t; W^-_t}$ is an infinite vector with entries of the covariances of $Y_t$ with $W_{t-1},\,W_{t-2},\ldots$ and 
\begin{align*}
\Gamma_{Y^-_t; W^-_t} =
\left(\begin{array}{cccc}
\Gamma_{YW}(0)            & \Gamma_{YW}(1)         & \ldots        \\
\Gamma_{WY}(1)    & \Gamma_{YW}(0)           & \ldots           \\
\vdots                   & \vdots                      & \ddots               \\
\end{array}
\right).
\end{align*}
The quotient in Eq.~(\ref{eq:ana_model_ent}) of these infinite dimensional matrices is difficult if not impossible to evaluate in the general case. 
Here, we will only consider two simple cases. 

\subsubsection{$c_{XZ}=c_{WY}=0$}
For the case of $c_{XZ}=c_{WY}=0$, i.e., as inputs solely an autodependency in $X$, the covariance matrix takes the simple form
\begin{align}
\Gamma_{Y_t Y^-_t W^-_t Z^-_t} &= 
\left(\begin{array}{cccc}
\Gamma_{Y_t}            & \Gamma_{Y_t; Y^-_t}        & 0        & 0 \\
\Gamma_{Y_t; Y^-_t}^\top & \Gamma_{Y^-_t }           & 0       & 0 \\
0                   & 0                   & \Gamma_{W^-_t}              & 0 \\
0                    & 0                      & 0      & \Gamma_{ Z^-_t} 
\end{array}
\right)
\end{align}
where the top left block is an infinite dimensional Toeplitz matrix, i.e., a Toeplitz operator. Then the quotient in Eq.~(\ref{eq:ana_model_ent}) can be simplified to 
\begin{align}
 \frac{|\Gamma_{Y_t Y^-_t}||\Gamma_{W^-_t Z^-_t}|}{|\Gamma_{Y^-_t}||\Gamma_{W^-_t Z^-_t}|} = \frac{|\Gamma_{Y_t Y^-_t}|}{|\Gamma_{Y^-_t}|}.
\end{align}
$\Gamma_{Y_t Y^-_t}$ and $\Gamma_{Y^-_t}$ are the symmetric Toeplitz matrices $G_{\tau}$ and $G_{\tau-1}$ with diagonal elements $\Gamma_{Y}$ and off-diagonal elements $g_{\tau}$ 
\begin{align}
g_0 = \Gamma_Y &= c_{XY}^2 \frac{\sigma_X^2}{1-a_X^2} +  \sigma_Y^2  &\\
g_{\tau} &= a_X^{|\tau|} \frac{c_{XY}^2 \sigma_X^2}{1-a_X^2} &\text{for}~~\tau \geq 1.
\end{align}
The desired TE is then given by
\begin{align}
I^{\rm TE}_{X\to Y} = \lim_{\tau\to \infty} \frac{1}{2} \ln \frac{1}{\sigma_Y^2} \frac{| G_\tau|}{|G_{\tau-1}|}.
\end{align}
To obtain the limit of the ratio of Toeplitz matrices we can utilize Szeg\"o's theorem \cite{szegoe,boettcher2006} which relates the limit to the geometric mean of a function $f(\lambda)$
\begin{align}
\lim_{\tau\to \infty} \frac{| G_\tau (f) |}{| G_{\tau-1} (f) |} = \exp \left( \frac{1}{2 \pi} \int_0^{2 \pi} \ln f(\lambda) d\lambda \right),
\end{align}
which requires that the Toeplitz matrix is in the Wiener class, i.e. the entries must be absolutely summable, which we assume here. 
The function $f(\lambda)$ is the Fourier series with the entries of the Toeplitz matrix being the coefficients
\begin{align}
&f(\lambda)= \sum_{\tau=-\infty}^{\infty} g_\tau e^{i \tau \lambda} = \Gamma_Y + 2 \sum _{\tau=1}^{\infty} g_\tau e^{i \tau \lambda} \\
&= c_{XY}^2 \frac{\sigma_X^2}{1-a_X^2} +  \sigma_Y^2 + 2 \frac{c_{XY}^2 \sigma_X^2}{1-a_X^2} \underbrace{\sum_{\tau=1}^{\infty} a_X^{|\tau|} e^{i \tau \lambda}}_{\frac{a_X e^{i \lambda} }{1- a_X e^{i \lambda}}}\\
&=\frac{\overbrace{[c_{XY}^2\sigma_X^2 - \sigma_Y^2(1{-}a_X^2)]a_X}^{\alpha}e^{i \lambda} + \overbrace{c_{XY}^2\sigma_X^2 + \sigma_Y^2(1{-}a_X^2)}^{\beta}}{(1-a_X^2) (1-a_X e^{i \lambda})}
\end{align}
with $\alpha < \beta$ for $|a_X|<1$.
Then the TE is
\begin{align}
I^{\rm TE}_{X\to Y} &= \lim_{\tau\to \infty} \frac{1}{2} \ln \frac{1}{\sigma_Y^2} \frac{| G_\tau|}{|G_{\tau-1}|} \nonumber \\
                    &= \lim_{\tau\to \infty} \frac{1}{2} \ln \frac{| G_\tau|}{|G_{\tau-1}|}   - \frac{1}{2} \ln \sigma_Y^2 \\
                    &=  \frac{1}{2} \ln \lim_{\tau\to \infty} \frac{| G_\tau|}{|G_{\tau-1}|}   - \frac{1}{2} \ln \sigma_Y^2 \\
                    &= \frac{1}{2} \ln \exp \left( \frac{1}{2 \pi} \int_0^{2 \pi} \ln f(\lambda) d\lambda \right)   - \frac{1}{2} \ln \sigma_Y^2 \\
                    &= \frac{1}{4 \pi} \int_0^{2 \pi} \ln f(\lambda) d\lambda  - \frac{1}{2} \ln \sigma_Y^2 \\
                    &= \frac{1}{4 \pi} \left[ \underbrace{\int_0^{2 \pi} \ln \left( \alpha e^{i \lambda} + \beta  \right) d\lambda}_{(\star)}  - \ln (1-a_X^2) \underbrace{\int_0^{2 \pi} d\lambda}_{2 \pi} \right. \nonumber \\
          &~~~~~ \left.  - \underbrace{\int_0^{2 \pi} \ln \left(  1-a_X e^{i \lambda}  \right) d\lambda}_{(\star\star)}    \right] - \frac{1}{2} \ln \sigma_Y^2,  
\end{align}
where the integrals $(\star)$ and $(\star\star)$ can be evaluated using contour integration to 
\begin{align}
(\star) &= 2\pi \ln \beta = 2\pi\ln \left( c_{XY}^2\sigma_X^2 + \sigma_Y^2(1{-}a_X^2) \right)   &\text{for $\alpha \leq \beta$},\\
(\star\star) &= 2\pi \ln  1 = 0   &\text{for $a_X \leq 1$}.
\end{align}
The TE is thus
\begin{align}
I^{\rm TE}_{X\to Y} &=\frac{1}{2}\ln \left( 1 + \frac{(c_{XY}^2 \sigma_X^2)/(1{-}a_X^2)}{\sigma_Y^2} \right)
\end{align}
and depends on the autodependency strength of $X$.

\subsubsection{$a_{X}=0$}
Now the process ``decouples in time'' since no autodependencies are present. The covariance matrix is
\begin{align}
\Gamma_{Y_t Y^-_t W^-_t Z^-_t} &= 
\left(\begin{array}{cccc}
\Gamma_{Y_t}            & 0        & \Gamma_{Y_t; W^-_t}         & \Gamma_{Y_t; Z^-_t} \\
0 & \Gamma_{Y^-_t }           & \Gamma_{Y^-_t; W^-_t}       & \Gamma_{Y^-_t; Z^-_t} \\
\Gamma_{Y_t; W^-_t}^\top & \Gamma_{Y^-_t; W^-_t}^\top & \Gamma_{W^-_t}              & 0 \\
\Gamma_{Y_t; Z^-_t}^\top & \Gamma_{Y^-_t; Z^-_t}^\top & 0     & \Gamma_{ Z^-_t} 
\end{array}
\right).
\end{align}
with the blocks being
\begin{align*}
\Gamma_{Y_t} &= c_{WY}^2 \sigma _W^2+c_{XY}^2 \sigma _X^2+\sigma _Y^2 \\
\Gamma_{Y_t; W^-_t}  &= (c_{WY} \sigma^2_W,0,0,\ldots) \\
\Gamma_{Y_t; Z^-_t}  &= (c_{XY} c_{XZ} \sigma_X^2,0,0,\ldots) \\
\Gamma_{Y^-_t}  &= (c_{WY}^2 \sigma _W^2+c_{XY}^2 \sigma _X^2+\sigma _Y^2) \mathbb{I} \\
\Gamma_{Y^-_t; W^-_t} &= c_{WY} \sigma^2_W \mathbb{S} \\
\Gamma_{Y^-_t; Z^-_t}  &=   c_{XY} c_{XZ} \sigma_X^2 \mathbb{S} \\
\Gamma_{W^-_t} &= \sigma_W^2 \mathbb{I} \\
\Gamma_{Z^-_t} &= ( c_{XZ}^2 \sigma_X^2+\sigma_Z^2) \mathbb{I}
\end{align*}
where $\mathbb{I}$ is the identity matrix and $\mathbb{S}$ is the shift matrix with ones on the superdiagonal, i.e., the first upper off-diagonal, and zeros everywhere else. 
The quotient in Eq.~(\ref{eq:ana_model_ent}) can be simplified by expressing the block matrix in terms of the Schur complement of the covariance block $\Gamma_{Y^-_t W^-_t Z^-_t}$
\begin{align}
&\frac{|\Gamma_{Y_t Y^-_t W^-_t Z^-_t}|}{|\Gamma_{Y^-_t W^-_t Z^-_t}|} = \nonumber\\
&\left|\Gamma_{Y_t} - (\Gamma_{Y_t; Y^-_t},\Gamma_{Y_t; W^-_t},\Gamma_{Y_t; Z^-_t}) (\Gamma_{Y^-_t W^-_t Z^-_t})^{-1} \left( \begin{smallmatrix} \Gamma_{Y_t; Y^-_t}^\top\\ \Gamma_{Y_t; W^-_t}^\top \\ \Gamma_{Y_t; Z^-_t}^\top \end{smallmatrix} \right)  \right|.
\end{align}
Since the vector $(\Gamma_{Y_t; Y^-_t},\Gamma_{Y_t; W^-_t},\Gamma_{Y_t; Z^-_t})$ contains only two non-zero elements, we do not have to take the infinite limit and do not need to invert the whole matrix $\Gamma_{Y^-_t W^-_t Z^-_t}$. A simple calculation yields
\begin{align}
&\frac{|\Gamma_{Y_t Y^-_t W^-_t Z^-_t}|}{|\Gamma_{Y^-_t W^-_t Z^-_t}|} = \nonumber \\
&= c_{WY}^2 \sigma _W^2+c_{XY}^2 \sigma _X^2+\sigma _Y^2 - \frac{c_{WY}^2 \sigma^4_W}{\sigma_W^2} - \frac{c_{XY}^2 c_{XZ}^2 \sigma_X^4}{c_{XZ}^2 \sigma_X^2+\sigma_Z^2},
\end{align}
from which we get
\begin{align}
I^{\rm TE}_{X\to Y} &= \frac{1}{2}\ln \left( 1 + \frac{c_{XY}^2 \sigma_X^2 \sigma_Z^2}{\sigma_Y^2(c_{XZ}^2\sigma_X^2+\sigma_Z^2)} \right).
\end{align}
Here, the TE depends on the coupling strength of $X$ with $Z$, which seems rather unintuitive.
This formula could have also been derived by exploiting separation properties of the corresponding time series graph (i.e., Markov properties of the process), from which a much smaller set of conditions can be inferred.

\subsection{MIT and related measures}
The measures based on the parental sets are much easier to derive because they involve only finite and very low dimensional covariance matrices.
As an example, for the entropy $H(Y_t~|~W_{t-1},X_{t-3})$ needed to compute the MIT, the covariance matrix of $(Y_t,W_{t-1},X_{t-3})$ is
\begin{align}
\left(
\begin{array}{ccc}
 c_{WY}^2 \sigma_W^2+\frac{c_{XY}^2 \sigma_X^2}{1{-}a_X^2}+\sigma_Y^2 & c_{WY} \sigma_W^2 & \frac{a_X c_{XY} \sigma_X^2}{1{-}a_X^2} \\
 c_{WY} \sigma_W^2 & \sigma_W^2 & 0 \\
 \frac{a_X c_{XY} \sigma_X^2}{1{-}a_X^2} & 0 & \frac{\sigma_X^2}{1{-}a_X^2} \\
\end{array}
\right).
\end{align}

\section{Proof of coupling strength autonomy theorem}
\noindent

To compute MIT,
\begin{align*}
I^{\rm MIT}_{X{\to}Y}(\tau) &\equiv I(X_{t-\tau};Y_t| \mathcal{P}_{Y_t} {\setminus}\{X_{t-\tau}\},\mathcal{P}_{X_{t-\tau}} )  \\
    &=  H(Y_t| \mathcal{P}_{Y_t} {\setminus}\{X_{t-\tau}\},\mathcal{P}_{X_{t-\tau}} ) - H(Y_t| \mathcal{P}_{Y_t} )
\end{align*}
 we need the source entropy $H(Y_t|\mathcal{P}_{Y_t})$ and the conditional entropy $H(Y_t|\mathcal{P}_{Y_t} {\setminus}\{X_{t-\tau}\},\mathcal{P}_{X_{t-\tau}})$. 
For the following steps we firstly use the independence of the i.i.d. variables $\eta^X_{t-\tau}$ and $\eta^Y_t$ of processes in the past, i.e., $I(\eta^Y_t;\mathbf{X}^-_t)=0$, and further due to the data processing inequality \cite{Cover2006} also 
\begin{align} \label{eq:indep}
I(\eta^Y_t;\tilde{f}(\mathbf{X}^-_t))=0
\end{align}
and correspondingly $I(\eta^X_{t-\tau};\tilde{g}(\mathbf{X}^-_{t-\tau}))=0$ for arbitrary functions $\tilde{f},\,\tilde{g}$. This implies in particular $I(\eta^Y_t;\tilde{f}(\mathcal{P}_{Y_t}))=0$ and $I(\eta^X_{t-\tau};\tilde{g}(\mathcal{P}_{X_{t-\tau}}))=0$.
Secondly, we use that generally for random variables $Y$ and $W$ and an arbitrary function $f$ 
\begin{align}
H(Y + f(W) | W ) &= \int p(w) H(Y + f(W) | W=w ) dw \nonumber\\
            & = \int p(w) H(Y | W=w ) dw \nonumber \\
            &= H(Y|W), \label{eq:trans_inv}
\end{align}
because $f(W)$ for $W=w$ is a fixed constant and entropies are translationally invariant.

Then, for $\tilde{f}(\mathcal{P}_{Y_t})=f(X_{t-\tau}) + g_Y(\mathcal{P}_{Y_t}{\setminus}\{X_{t-\tau}\})$, the source entropy is
\begin{align}
H(Y_t|\mathcal{P}_{Y_t}) &= H(\tilde{f}(\mathcal{P}_{Y_{t}}) +  \eta^Y_t|\mathcal{P}_{Y_t}) \\
                         &= H(\eta^Y_t|\mathcal{P}_{Y_t}) \\
                         &= H(\eta^Y_t),
\end{align}
and depends only on the distribution of the source process $\eta^Y_t$. This relation holds generally if $Y_t$ additively depends on its parents.

Next, to compute the other conditional entropy, we insert Eq.~(\ref{eq:sepa_x}) in (\ref{eq:sepa_y}) and get
\begin{align} 
&H(f(\eta^X_{t-\tau}+g_X(\mathcal{P}_{X_{t-\tau}})) + g_Y(\mathcal{P}_{Y_t} {\setminus}\{X_{t-\tau}\}) + \eta^Y_t | \nonumber\\
&~~~~~~~~~~~~~~~~~~~~~~~~~~~~~~~~~~~~~~~~~~~~~~~~~~~~~| \mathcal{P}_{Y_t} {\setminus}\{X_{t-\tau}\},\mathcal{P}_{X_{t-\tau}}) \nonumber \\ 
&=H(f(\eta^X_{t-\tau}+g_X(\mathcal{P}_{X_{t-\tau}})) +  \eta^Y_t | \mathcal{P}_{Y_t} {\setminus}\{X_{t-\tau}\},\mathcal{P}_{X_{t-\tau}}) \label{eq:ent_cond1}
\end{align}
also due to translational invariance. If we only assume condition (1) this relation cannot be much further simplified.

To arrive at a CMI again, we need to expand the source entropy using Eq.~(\ref{eq:trans_inv}) and (\ref{eq:indep}).
First, we add the same conditions as in Eq.~(\ref{eq:ent_cond1}), which is possible since $\eta^Y_t$ is independent of all past processes:
\begin{align}
H(\eta^Y_t) = H(\eta^Y_t ~|~\mathcal{P}_{Y_t} {\setminus}\{X_{t-\tau}\},\mathcal{P}_{X_{t-\tau}}).
\end{align}
Next, we insert the term $f(\eta^X_{t-\tau}+g_X(\mathcal{P}_{X_{t-\tau}}))$ and ``condition it out again'' using Eq.~(\ref{eq:trans_inv}) by adding $\eta^X_{t-\tau}$ to the conditions ($\mathcal{P}_{X_{t-\tau}}$ is already included):
\begin{align}
H(\eta^Y_t) &= H(\eta^Y_t ~|~\mathcal{P}_{Y_t} {\setminus}\{X_{t-\tau}\},\mathcal{P}_{X_{t-\tau}}) \nonumber \\
&=H(f(\eta^X_{t-\tau}+g_X(\mathcal{P}_{X_{t-\tau}})) + \eta^Y_t ~|~\nonumber\\
&~~~~~~~~~~~~~~~~~~|~\mathcal{P}_{Y_t} {\setminus}\{X_{t-\tau}\},\mathcal{P}_{X_{t-\tau}},\eta^X_{t-\tau}).
\end{align}
Then via
\begin{align*}
&I^{\rm MIT}_{X{\to}Y}(\tau) = \nonumber\\
& H(f(\eta^X_{t-\tau}+g_X(\mathcal{P}_{X_{t-\tau}})) +  \eta^Y_t ~|~ \mathcal{P}_{Y_t} {\setminus}\{X_{t-\tau}\},\mathcal{P}_{X_{t-\tau}}) -\nonumber\\
&H(f(\eta^X_{t-\tau}+g_X(\mathcal{P}_{X_{t-\tau}})) + \eta^Y_t ~|~\mathcal{P}_{Y_t} {\setminus}\{X_{t-\tau}\},\mathcal{P}_{X_{t-\tau}},\eta^X_{t-\tau})
\end{align*}
we arrive at Eq.~(\ref{eq:mit_3}). 

If we assume conditions~(1) and (2), we can further simplify Eq.~(\ref{eq:ent_cond1}) since $f(\eta^X_{t-\tau}+g_X(\mathcal{P}_{X_{t-\tau}}))=c\eta^X_{t-\tau}+c g_X(\mathcal{P}_{X_{t-\tau}})$ and therefore
\begin{align} 
&H(c\eta^X_{t-\tau}+c g_X(\mathcal{P}_{X_{t-\tau}}) +  \eta^Y_t | \mathcal{P}_{Y_t} {\setminus}\{X_{t-\tau}\},\mathcal{P}_{X_{t-\tau}}) \nonumber\\
&=H(c\eta^X_{t-\tau}+  \eta^Y_t | \mathcal{P}_{Y_t} {\setminus}\{X_{t-\tau}\})
\end{align}
where we used Eq.~(\ref{eq:trans_inv}) and the fact that $I(c\eta^X_{t-\tau}+  \eta^Y_t;\mathcal{P}_{X_{t-\tau}}~|~\mathcal{P}_{Y_t} {\setminus}\{X_{t-\tau}\})=0$ (also holds without the condition on $\mathcal{P}_{Y_t} {\setminus}\{X_{t-\tau}\}$ because $\mathcal{P}_{X_{t-\tau}}$ lies in the past of both $\eta^X_{t-\tau}$ and $\eta^Y_t$). Extending the source entropy again we arrive at Eq.~(\ref{eq:mit_2}). If the ``sidepath''-parents in 
\begin{align} \label{eq:sidepath_parents}
\mathcal{P}^\star_{Y_t} \equiv \{W^k_{t-\tau_k}\in \mathcal{P}_{Y_t}{\setminus}\mathcal{P}_{X_{t-\tau}}:~~ I(\eta^X_{t-\tau}; W^k_{t-\tau_k})>0\}
\end{align}
are additively separated from the remaining parents, MIT can be further simplified.

If additionally condition (3) holds, then Eq.~(\ref{eq:nosidepath}) leads to $I(c\eta^X_t+  \eta^Y_t;\mathcal{P}_{Y_{t}}{\setminus}\{X_{t-\tau}\})=0$, and we, therefore, can drop $\mathcal{P}_{Y_t} {\setminus}\{X_{t-\tau}\}$ from the conditions from which Eq.~(\ref{eq:mit_1}) follows. 

For the contemporaneous MIT 
\begin{align*} 
 I^{\rm MIT}_{X{\--}Y} &\equiv I(X_{t};Y_t| \mathcal{P}_{Y_t},\mathcal{P}_{X_{t}},\mathcal{N}_{X_t}{\setminus}\{Y_t\},\mathcal{N}_{Y_t}{\setminus}\{X_t\},\\
&~~~~~~~~~~~~~~~~~~~~~\mathcal{P}(\mathcal{N}_{X_t}{\setminus}\{Y_t\}),\mathcal{P}(\mathcal{N}_{Y_t}{\setminus}\{X_t\}) )  
\end{align*}
we only need condition~(1) for which the entropy in the first term
\begin{align}
&H(\eta^Y_t+g_Y(\mathcal{P}_{Y_t})| \mathcal{P}_{Y_t},\mathcal{P}_{X_{t}},\mathcal{N}_{X_t}{\setminus}\{Y_t\},\mathcal{N}_{Y_t}{\setminus}\{X_t\},\nonumber\\
&~~~~~~~~~~~~~~~~~~~~~\mathcal{P}(\mathcal{N}_{X_t}{\setminus}\{Y_t\}),\mathcal{P}(\mathcal{N}_{Y_t}{\setminus}\{X_t\}) ) \\
&=H(\eta^Y_t| \mathcal{P}_{Y_t},\mathcal{P}_{X_{t}},\mathcal{N}_{X_t}{\setminus}\{Y_t\},\mathcal{N}_{Y_t}{\setminus}\{X_t\},\nonumber\\
&~~~~~~~~~~~~~~~~~~~~~\mathcal{P}(\mathcal{N}_{X_t}{\setminus}\{Y_t\}),\mathcal{P}(\mathcal{N}_{Y_t}{\setminus}\{X_t\}) ) \\
&=H(\eta^Y_t| \mathcal{N}_{X_t}{\setminus}\{Y_t\},\mathcal{N}_{Y_t}{\setminus}\{X_t\}),
\end{align}
again due to translational invariance of entropy [Eq.~(\ref{eq:trans_inv})] and the independence of $\eta^Y_t$ of past processes [Eq.~(\ref{eq:indep})]. For the same reasons the entropy in the second term becomes
\begin{align}
&H(\eta^Y_t+g_Y(\mathcal{P}_{Y_t})| \mathcal{P}_{Y_t},\mathcal{P}_{X_{t}},\mathcal{N}_{X_t}{\setminus}\{Y_t\},\mathcal{N}_{Y_t}{\setminus}\{X_t\},\nonumber\\
&~~~~~~~~~~~~~~~~~~~~~\mathcal{P}(\mathcal{N}_{X_t}{\setminus}\{Y_t\}),\mathcal{P}(\mathcal{N}_{Y_t}{\setminus}\{X_t\}),X_t ) \nonumber\\
&=H(\eta^Y_t| \mathcal{P}_{Y_t},\mathcal{P}_{X_{t}},\mathcal{N}_{X_t}{\setminus}\{Y_t\},\mathcal{N}_{Y_t}{\setminus}\{X_t\},\nonumber\\
&~~~~~~~~~~~~~~~~~~~~~\mathcal{P}(\mathcal{N}_{X_t}{\setminus}\{Y_t\}),\mathcal{P}(\mathcal{N}_{Y_t}{\setminus}\{X_t\}),\eta^X_t+g_X(\mathcal{P}_{X_t}) ) \nonumber \\
&=H(\eta^Y_t| \mathcal{N}_{X_t}{\setminus}\{Y_t\},\mathcal{N}_{Y_t}{\setminus}\{X_t\},\eta^X_t),
\end{align}
because knowing $\eta^X_t+g_X(\mathcal{P}_{X_{t}})$ and $\mathcal{P}_{X_{t}}$ is equivalent to knowing $\eta^X_t$ and $\mathcal{P}_{X_{t}}$. Then Eq.~(\ref{eq:mit_contemp}) follows which finishes the proof.

Similarly, MITS and MITN can be simplified if the dependency $g_Y$ is additive in the parents.

\section{Further Numerical Experiments}
\begin{figure}[t]
\includegraphics[width=\columnwidth]{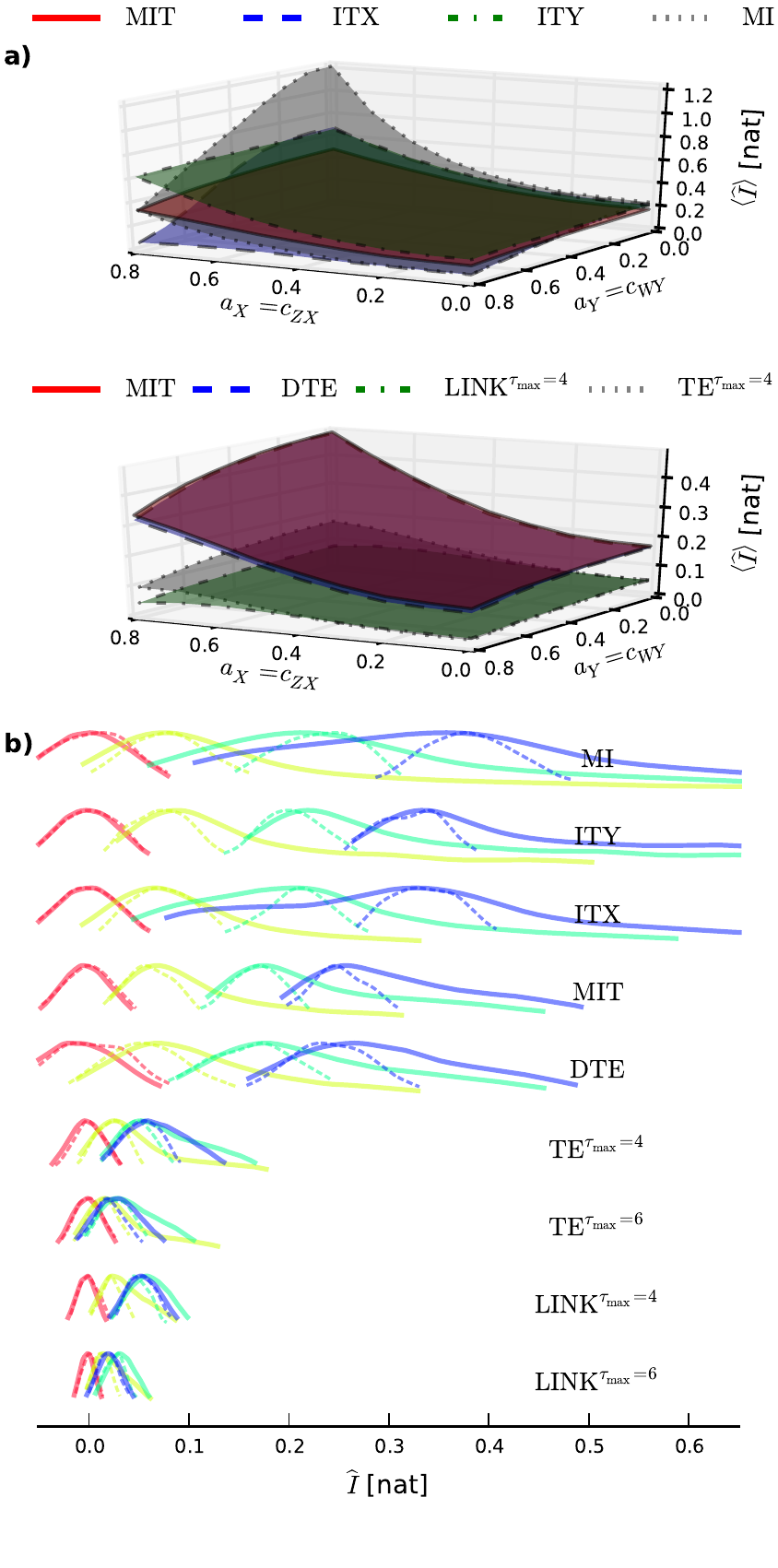}
\caption[]{(Color online) Numerical experiments with the model Eq.~(\ref{eq:model}) with setup as before but for squared dependency $f(x)=x^2$ with $c_{XY}=0.6$.
}
\label{fig:numerics}
\end{figure}
In Fig.~\ref{fig:numerics} we show results of our numerical experiments for the model class Eq.~(\ref{eq:model}) with a nonlinear dependency  $f(x)=x^2$ of the link ``$X_{t-2}\to~Y_t$'' using the same ensemble setup $E$ as before. As discussed in Sect.~\ref{sect:analytics}, then the source process $\eta^X_{t-\tau}$ mixes with its parents and it does not make sense to attribute the coupling strength to one single coefficient.
As a result, the average of MIT in Fig.~\ref{fig:numerics}(a) tends to larger values for increased $a_X=c_{ZX}$, thus the inputs are not entirely ``filtered out''. Still, MIT is much less affected than MI. 

Regarding the inequality relation Eq.~(\ref{eq:inequality}), a nonlinear dependency does not affect at least the right side $I^{\rm MIT}_{X{\to}Y}(\tau)\leq I^{\rm ITY}_{X{\to}Y}(\tau)$ as demonstrated in Fig.~\ref{fig:numerics}(a) and (b). Although the left side of the inequality relation $I^{\rm ITX}_{X{\to}Y}(\tau)\leq I^{\rm MIT}_{X{\to}Y}(\tau)$ should hold under the same general condition\,(S) in \cite{Eichler2011} and the ``no sidepath''-constraint, it seems to be violated for large $a_X=c_{ZX}$ (and small $a_Y=c_{WY}$). This could be related to highly skewed distributions for nonlinear $f(x)$.

In the bottom plot of Fig.~\ref{fig:numerics}(a) it might seem, that TE and LINK are less affected, but actually the relative variance is much higher.



\begin{thebibliography}{10}


\bibitem{science2011}
Science Staff, Science {\bf 331},  692  (2011).

\bibitem{Reshef2011}
D.~N. Reshef, Y.~N. Reshef, H.~K. Finucane, S.~R. Grossman, G. McVean, P.~J. Turnbaugh, E.~S. Lander, M. Mitzenmacher, and P.~C. Sabeti, Science {\bf 334},  1518  (2011).

\bibitem{Schreiber2000}
T. Schreiber, Phys. Rev. Lett. {\bf 85},  461  (2000).

\bibitem{Cover2006}
T. Cover and J. Thomas, {\em {Elements of Information Theory}} (John Wiley \&  Sons, New York, 2006).

\bibitem{Schweizer1981}
B. Schweizer and E.~F. Wolff, Ann. Stat. {\bf 9}(4), 879 (1981)

\bibitem{Gorfine2012}
M. Gorfine, R. Heller, and Y. Heller, {\tt http://iew3.technion.ac.il/gorfinm/files/\\science6.pdf}.

\bibitem{Jachan2009a}
M. Jachan, K. Henschel, J. Nawrath, A. Schad, J. Timmer, and B. Schelter, Phys. Rev. E {\bf 80},  011138  (2009).

\bibitem{Schelter2009}
B. Schelter, J. Timmer, and M. Eichler, J. Neuroscience Methods {\bf 179},  121  (2009).

\bibitem{Chen2004}
Y. Chen, G. Rangarajan, J. Feng, and M. Ding, Phys. Lett. A {\bf 324}, 26  (2004).

\bibitem{Marinazzo2008}
D. Marinazzo, M. Pellicoro, and S. Stramaglia, Phys. Rev. Lett. {\bf 100}, 144103  (2008).

\bibitem{Barnett2009}
﻿L. Barnett, A.~B. Barrett, and A.~K. Seth, Phys. Rev. Lett. {\bf 103}, 238701  (2009).

\bibitem{Faes2011}
L. Faes, G. Nollo, and A. Porta, Phys. Rev. E {\bf 83}, 051112 (2011).

\bibitem{Runge2012prl}
J. Runge, J. Heitzig, V. Petoukhov, and J. Kurths, Phys. Rev. Lett. {\bf  108}, 258701 (2012).

\bibitem{ay2008information}
N. Ay and D. Polani, Adv. in Compl. Sys. {\bf 11},  17  (2008).

\bibitem{Janzing2012}
D. Janzing, D. Balduzzi, M. Grosse-Wentrup, and B. Schoelkopf,  {\tt  	arXiv:1203.6502v1 [math.ST]}    (2012).

\bibitem{Shannon1963}
C. Shannon and W. Weaver, {\em The Mathematical Theory of Communication}  (University of Illinois Press, Urbana, 1963).

\bibitem{Pompe2011}
B. Pompe and J. Runge, Phys. Rev. E {\bf 83},  051122  (2011).

\bibitem{lauritzen1996graphical}
S.~L. Lauritzen, {\em {Graphical Models}}, Oxford Statistical Science Series, Vol. 16 (Clarendon,  Oxford, 1996).

\bibitem{Dahlhaus2000}
R. Dahlhaus, Metrika {\bf 51},  157  (2000).

\bibitem{Eichler2011}
M. Eichler, Prob. Theo. and Rel. Fields {\bf 1}, 233 (2012).

\bibitem{szegoe}
﻿G. Szeg\"o, Math. Annalen {\bf 76}, 4 (1915).

\bibitem{boettcher2006}
A. ﻿B\"ottcher, and B. Silbermann, {\em { Analysis of Toeplitz Operators}} (Springer Verlag, New York, 2006).

\bibitem{brockwell2009time}
P. Brockwell and R. Davis, {\em {Time Series: Theory and Methods}} (Springer Verlag, New York, 2009).

\bibitem{Pearl2000}
J. Pearl, {\em { Causality: Models, Reasoning, and Inference}} (Cambridge Univ. Press, Cambridge, 2000).

\bibitem{Kraskov2004a}
A. Kraskov, H. St\"ogbauer, and P. Grassberger, Phys. Rev. E {\bf 69},  066138  (2004).

\bibitem{FrenzelPompe2007}
S. Frenzel and B. Pompe, Phys. Rev. Lett. {\bf 99},  204101   (2007).

\bibitem{Compo2006}
G. Compo, J. Whitaker, and P. Sardeshmukh, Bull. of the Am.  Met. Soc. {\bf 87},  175  (2006).

\bibitem{Lau2002}
K. Lau and S. Yang, Encyclopedia of Atmospheric Sciences,  2505  (2003).

\end{thebibliography}
\end{document}